\documentclass[prb,aps,amsmath,amssymb,showpacs,twocolumn,longbibliography,nofootinbib]{revtex4-2}

\usepackage{amsmath,amssymb,amsfonts,graphicx,float,dcolumn,bm,color,colortbl,multirow}
\usepackage[titletoc,title]{appendix}
\usepackage[dotinlabels]{titletoc}
\usepackage{cancel}
\usepackage{braket}
\usepackage{mathrsfs}
\usepackage[mathscr]{euscript}
\usepackage{comment}

\usepackage[dvipsnames]{xcolor}

\newcommand{\quotes}[1]{``#1''}

\usepackage[colorlinks=true, citecolor=blue, linkcolor=blue, urlcolor=blue]{hyperref}
\usepackage{cleveref}

\begin{document}
	
	% Title section
	\title{Signatures of quantum chaos and complexity in the Ising model on random graphs}

        \author{G. J. Sreejith}
        \affiliation{Indian Institute of Science Education and Research, Pune 411008, India}

        \author{Sandipan Manna}
        \thanks{Corresponding author}
        \email{sandipan.manna@students.iiserpune.ac.in}
        \affiliation{Indian Institute of Science Education and Research, Pune 411008, India}
        
	\begin{abstract}
We investigate signatures of quantum chaos in the mixed-field quantum Ising model on finite-size Erdős-Rényi graphs using probes scalable on near-term quantum devices. Upon tuning the graph connectivity, the system exhibits a crossover from a localized regime at low connectivity, through a chaotic regime at intermediate connectivity, to a permutation-symmetric integrable limit near all-to-all connectivity. This crossover has possible implications for the performance and trainability of variational algorithms such as QAOA.
We characterize this crossover in finite-size systems using complementary probes. First, deep thermalization of a projected ensemble starting from a product state reveals slow (fast) convergence to the Haar ensemble at extremal (intermediate) connectivities.
Second, we analyze eigenstate and eigenvalue correlations using the partial spectral form factor, an experimentally scalable proxy for the spectral form factor with reduced resource overhead, and observe characteristic chaos signatures at intermediate connectivities and distinct deviations at extremal connectivities. Finally, we explore the Krylov complexity of operators, a locality-independent diagnostic that, although not directly experimentally accessible, serves as a tool for quantifying scrambling. We show that it is maximized deep in the chaotic regime, corroborating the signatures observed through the experimentally scalable probes. Our results provide finite-size benchmarks demonstrating robust signatures of chaos in scalable probes and suggest that these diagnostics can be implemented in current quantum platforms to access regimes beyond classical simulation.
	\end{abstract}
	\maketitle
	
	% Main content
	\section{Introduction}
Quantum Ising models have played a central role in the many-body physics of quantum phase transitions~\cite{sachdev2023quantum,blote2002cluster, Dutta_Aeppli_Chakrabarti_Divakaran_Rosenbaum_Sen_2015}, arising naturally in lattice systems, disordered magnets, and glassy settings ~\cite{PhysRevLett.35.1792,RevModPhys.58.801,santoro2002theory}. 
They also serve as the framework for quantum annealing (QAA) and variational quantum algorithms (VQA), which encode classical optimization problems into an Ising Hamiltonian and leverage quantum fluctuations to explore the energy landscape in search of low-energy solutions~\cite{albash2018adiabatic,zhou2020quantum,lucas2014ising,kadowaki1998quantum,Edward_QAA,RevModPhys.80.1061,hauke2020perspectives}. 
Hardware connectivity constraints~\cite{harrigan2021quantum,krantz2019quantum} as well as the need to encode general graph optimization problems~\cite{nguyen2023quantum,herrman2021impact,PhysRevApplied.5.034007} make the dynamics of these models on random graphs of direct practical relevance. 
Performance of related optimization circuits~\cite{cain2023quantum,Schl_mer_2025,moein_reservoir} has been shown, in specific instances, to correlate with chaos diagnostics such as operator spreading ~\cite{PhysRevLett.97.050401,vsuntajs2020quantum,braun2020transition} and information scrambling~\cite{Yasuhiro_Sekino_2008,Maldacena_2016,huang2017out,iyoda2018scrambling,tian2022testing,munoz2024quantum,mi2021information}. 
While these diagnostics have been investigated extensively in regular lattice systems~\cite{braun2020transition,huang2017out,PhysRevB.104.214202,PhysRevB.107.L220203,PhysRevX.8.031057,PhysRevLett.123.165902}, studies on irregular graphs remain largely unexplored. 
 
In QAA, the instantaneous Hamiltonian interpolates between a transverse-field driver and an Ising problem Hamiltonian encoded on a random graph. For initial states with finite excitation energy density, this interpolating Hamiltonian can exhibit chaotic dynamics. While ideal adiabatic evolution remains confined to the low-energy sector and is therefore insensitive to bulk spectral statistics, it has been shown that adding a chaotic perturbation at intermediate times can improve QAA performance in practical settings by enhancing the bottleneck gap and delocalizing low-energy states in hard optimization instances~\cite{Schl_mer_2025,cain2023quantum}. 

An alternative optimization scheme is the quantum approximate optimization algorithm (QAOA)~\cite{farhi2014quantum,zhou2020quantum}, which resembles a Trotterization of the QAA. The state is evolved under an alternating sequence of transverse field ($H_{\rm x}$) and cost ($H_{\rm c}$) Hamiltonians, with $\{\gamma_k, \beta_k\}$ as variational parameters, optimized to minimize $\langle H_{\rm c} \rangle$ on the final state.  Unlike in QAA, the intermediate states of a QAOA circuit are not constrained to the low-energy sector. Consequently, eigenstates drawn from the bulk of the spectrum of $H_{\rm c}$ or $H_{\rm x}$  may be transiently populated, making bulk spectral properties directly relevant to the circuit dynamics.  In simulations of randomly picked finite-size problem instances, we show (Sec.~\ref{QAA_section}) that a three-term ansatz with $H_{\rm c}$, $H_{\rm x}$, and an additional term $H_{\rm d}$ achieves a better approximation ratio than the standard two-term QAOA when $H_{\rm d}$ exhibits signatures of chaos. However, the same dynamics may also suppress parameter gradients at shallow depth, a precursor to barren plateaus~\cite{kim2023quantum,mcclean2018barren,wiersema2020exploring}. Quantum chaos has also been identified as a key resource in quantum reservoir computing. In a random-graph Ising setup closely related to the one studied here, it was found that optimal learning performance of a quantum reservoir is achieved at the boundary between the chaotic and localized phases~\cite{moein_reservoir}. Refined understanding and control of the onset of chaos in finite-size random-graph Hamiltonians can thus provide valuable insights into a range of proposed use cases for NISQ devices. 

In this work, we present an investigation of quantum dynamics in the Ising model on Erdős-Rényi (ER) graphs with a mixed field, focusing on how graph connectivity shapes the onset of quantum chaos. We parametrize graph connectivity using the connectance ($\tilde{M}$), the fraction of the maximum possible edges present in the graph. For simplicity, we consider the case in which Ising coupling strengths are held fixed; disorder in our model arises from the connectivity pattern and not from randomness in the coupling strengths or the external field.
Level spacing distributions, level velocity, and spectral correlations in this model were studied in Ref.~\cite{PhysRevResearch.7.013146}, where chaotic spectral properties were found to emerge in the intermediate $\tilde{M}$ regime. These are spectral diagnostics, as they characterize quantum chaos through eigenvalue correlations alone, and carry no direct information about eigenvector structure. Yet dynamical processes that are experimentally accessible, such as information scrambling, operator growth, and thermalization, are sensitive to eigenvector structure. To build a more complete picture, we employ diagnostics that probe signatures of chaos in projective measurement outcomes, eigenstate structure, operator spreading, and the growth of complexity under unitary evolution, by studying the projected ensemble statistics, partial spectral form factor, and Krylov complexity. These quantities are more directly connected to dynamical observables and, with the exception of Krylov complexity, to experimentally accessible measurements. Together, these provide a distinct and complementary perspective on quantum chaos.

Conventional studies of quantum thermalization focus on quantities governed by the eigenstate thermalization hypothesis (ETH): specifically, expectation values of local observables~\cite{deutsch1991quantum,srednicki1994chaos,rigol2008thermalization}. The projected ensemble (PE)~\cite{cotler2023emergent,choi2023preparing} offers a finer characterization by unraveling the density matrix into an ensemble of pure quantum states, enabling access to the full distribution of measurement outcomes beyond what low-order 
correlations capture. For systems without conserved charges, the PE of a time-evolved state is expected to converge to the Haar ensemble. This is known as \textit{deep thermalization}. We show that the distance between the PE and the Haar distribution decays faster with system size at intermediate $\tilde{M}$. In contrast, in sparse and dense limits ($\tilde{M} \to 0, 1$), the trace distance shows a slow decay with the system size. 
Next, we focus on the partial spectral form factor (pSFF), which extends the spectral form factor (SFF) to capture correlations between eigenstates and eigenvalues within the subsystem. We empirically demonstrate distinct signatures in the pSFF across different connectivity regimes. Measurements of PE and pSFF have already been experimentally demonstrated in $5-25$ qubit systems in various platforms~\cite{joshi2022probing,dong2025measuring,zhang2025holographic,choi2023preparing,yan2025characterizing}. While our numerical studies are restricted to $15$ qubits, near-term quantum devices can be used to extend this investigation to larger systems.

Heisenberg evolution under a chaotic Hamiltonian drives rapid growth of local operators, both in real space and in operator space~\cite{parker2019universal}. To probe the crossover from an operator-spreading perspective, we compute the Krylov complexity (KC) and its late-time saturation as a function of $\tilde{M}$. In the Krylov subspace, an initial local operator spreads over a set of orthogonal basis operators generated by Lanczos tridiagonalization of the Liouvillian superoperator, analogous to a particle propagating on a tight-binding chain with nearest-neighbor hopping. KC quantifies this by measuring the average position of the wave packet in the chain.  We observe that the Lanczos coefficients, which correspond to the hopping parameters in the tight-binding interpretation, exhibit enhanced fluctuations near the integrable and localized limits (at finite system sizes). Most importantly, the KC saturation values are significantly larger (and grow faster with system size) in the chaotic regime compared to the localized and integrable regimes. 

The crossover from chaotic to integrable and localized regimes is associated with the emergence of approximate conservation laws and local integrals of motion~\cite{Mierzejewski_2015,PhysRevB.91.085425,10.21468/SciPostPhys.14.5.125}. In the model studied here, this takes two distinct forms: at $\tilde{M} \to 0$, disconnected clusters yield trivial local conserved charges, while at $\tilde{M} = 1$, a permutation symmetry fragments the Hilbert space (Sec.~\ref{QAA_section}). At other values of $\tilde{M}$, this fragmentation is weakly broken, and the resulting approximate permutation symmetries act as quasi-conserved quantities that partially obstruct thermalization.

All three diagnostics used in this work, i.e., the projected ensemble, the partial spectral form factor, and the Krylov complexity, probe how these conservation laws suppress quantum information scrambling from complementary perspectives. PE deviates from the Haar ensemble in the presence of conserved quantities, instead approaching an ensemble constrained by the conserved charges. This was demonstrated in the case of $U(1)$ charges~\cite{mark2024maximum,manna2025projected} and kinetic constraints~\cite {bhore2023deep}. The universal ensemble approached under dynamics with more complex or approximate charges has not been studied extensively, and is expected to deviate from the Haar ensemble. The pSFF, an experimentally accessible proxy for the SFF, captures the effect of the crossover in eigenvalue and eigenstate correlations~\cite{Bohigas_1984,Berry_1977}. Krylov complexity tracks operator growth in Krylov space, with lower saturation values in integrable and localized settings reflecting restricted operator spreading. A central finding of this work is that all three consistently identify the same connectivity-driven crossover in finite-size systems, providing a unified picture of how approximate conservation laws govern the onset of chaos across these complementary facets of quantum dynamics.

This article is organized as follows. Section~\ref{QAA_section} provides a brief introduction to the model and its symmetries. We also briefly discuss the QAOA protocol and demonstrate the performance of random graph Ising model as a component in circuit ansatz. We utilize the projected ensemble protocol to examine deep thermalization through higher-order moments of quantum state ensembles in Sec.~\ref{PE_section}. In Sec.~\ref {pSFF_section}, we study the partial spectral form factor (pSFF) of subsystems to identify characteristic signatures of chaotic behavior and its implications in the context of experiments. We also explore the effect of the choice of subsystem on pSFF. Section~\ref{Krylov_section} presents results on operator delocalization through Krylov complexity. Finally, Sec.~\ref{Discussion_section} presents the summary and discusses the limitations and potential extensions of our study. 
     
	\section{Ising Model on random graphs	\label{QAA_section}
}	
	
	In this work, we consider the mixed-field quantum Ising model whose Hamiltonian is given by
    \begin{gather}
    \label{Hamiltonian}
    H = \frac{J}{M}H_{\rm p} + \frac{g}{L}H_{\rm x} + \frac{h}{L}H_{\rm z} \ , \\
    H_{\rm p} = -\sum_{i,j}A_{ij} Z_iZ_j, \quad H_{\rm x} = -\sum_i X_i, \quad H_{\rm z} = -\sum_i Z_i   \ ~.\nonumber
    \end{gather}
	$\{X_i, Y_i, Z_i\}$ are Pauli matrices acting on site $i$ and $g,h$ are the strengths of the transverse and longitudinal field, respectively. $A_{ij}$ is the adjacency matrix of the underlying graph, where spins are the nodes. $A_{ij}=1(0)$ if there is a coupling (no coupling) between spins $i$ and $j$. We take $J=1$ throughout this work. The adjacency matrices are chosen to represent random ER graphs~\cite{erdds1959random} $\mathcal{G}(L, M)$ where $L$ and $M$ denote the number of nodes and edges, respectively. $M$ can be obtained as $M=\frac{1}{2}\sum_{i,j}A_{ij}$. The denominators $L, M$ in Eq.~\ref{Hamiltonian} ensure that the bandwidth of the energy spectrum scales as $\approx O(JL^0M^0)$. At $g =  0$, the model is classical, with on-site $U(1)$ symmetries ($[H, Z_i] = 0$). For generic $g$, at  $h = 0$, the model has a global parity symmetry given by, $\mathcal{P}=\prod_{i=1}^L X_i $. All edges have equal strength ($J$) and graph-to-graph variation arises entirely from the adjacency matrix. For $L$ sites, the maximum number of edges is $M_{\text{max}} = \binom{L}{2}$. We define the connectance $\tilde{M} = M/M_{\text{max}}$ to characterize the graphs. For each $\tilde M$ we sample graphs with $\tilde{M}M_{\rm max}$ edges with uniform probability.
        \begin{figure}[htbp]
		\includegraphics[width=\columnwidth]{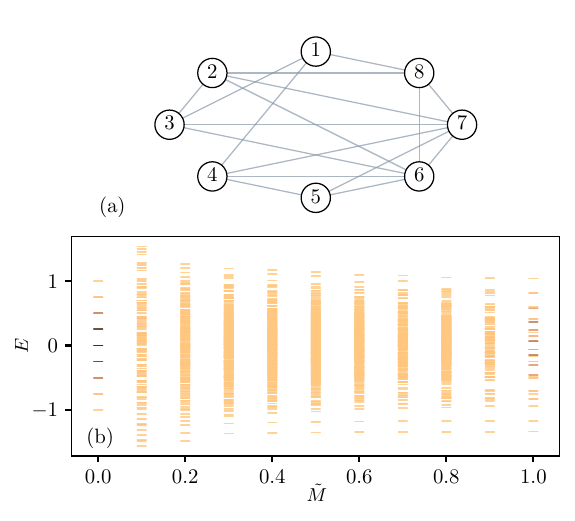}
		\caption{(a) A single realization of Erdős-Rényi graph with $L=8$ nodes and connectance $\tilde{M}=0.6$ which corresponds to number of connected edges, $M=17$. Interacting pairs are represented by connected edges. (b) Spectrum of Hamiltonian in Eq.~\ref{Hamiltonian} for an instance at each $\tilde{M}$ with $L=8$. We have used darker shades to indicate degeneracy.\label{erdos_graph}}
    \end{figure}
    
    $\tilde{M}=0$ and $1$ correspond to the non-interacting and integrable LMG~\cite{lipkin1965validity} limits, respectively (the presence of a small longitudinal field breaks the parity symmetry of the LMG model, while retaining integrability). Figure~\ref{erdos_graph}(a) shows one realization of such a graph. We take $ J=1$ and $ g=1$, thus working with a ferromagnetic Hamiltonian. $h$ is taken to be a small positive value in subsequent sections to break the model's parity symmetry. The dynamical properties studied in this work characterize the bulk of the spectrum, which is expected to be independent of the ferromagnetic or anti-ferromagnetic nature of the ground state, i.e., the sign of $J$.
    
The onset of chaos with $\tilde{M}$ in this model (with $h=0$) was studied  for finite systems in Ref.~\cite{PhysRevResearch.7.013146}  using level spacing statistics, level velocity, and spectral form factor. The system exhibits chaotic behavior at intermediate connectivity, with spectral features matching Wigner-Dyson (WD) statistics, and shows deviations at extreme $\tilde{M}$ in finite-system studies. Figure~\ref{erdos_graph}(b) shows the eigenvalue spectrum of this Hamiltonian at $L=8,g=1.0$ and $h=0.1$ for a representative realization at each $\tilde{M}$. For intermediate $\tilde M$, the level-spacing distribution in the unfolded spectrum is close to the WD distribution for numerically accessible systems. It approaches the WD distribution with increasing $L$ for $\tilde{M}$ close to $1$ and $0$ when $g$ takes a value in  $0.5J < g < 4J$~\cite{PhysRevResearch.7.013146}. Thus, at large enough $L$, and intermediate values of $g$, the level spacing ratio is expected to be WD for all $\tilde{M} \neq 0,1$.  At very large and small values of $g$, the system is nearly non-interacting, with eigenstates given by simple product states to a good approximation; thus, the level-spacing ratio deviates from WD. 

We choose $g=1$ throughout to retain the possibility of chaotic behavior in the system sizes studied ($6 \leq L \leq 15$). Next, we discuss the symmetries of this Hamiltonian at extremal $\tilde{M}$.

\subsection{Symmetries and conservation laws near $\tilde M=0$ and $1$}
At small $\tilde{M}$ and finite system sizes, ER graphs consist of weakly connected clusters. The Hamiltonian then decomposes effectively into smaller blocks with support only within individual clusters, leading to a breakdown of ergodicity. Consequently, eigenstates in this regime are weakly entangled, with bipartite entanglement entropy across a real-space cut remaining well below the volume-law value expected for chaotic systems, a trend we confirm for $\tilde{M} < 0.3$ (Fig.~\ref{entanglement_eigesnstate} in Appendix~\ref{ipr_appendix}). However, the percolation threshold for ER graphs is $\tilde{M}_c \sim 1/L$, so for any fixed small $\tilde{M} > 0$, increasing the system size $L$ eventually drives the graph past this threshold, beyond which a giant component of size $\mathcal{O}(L)$ emerges~\cite{erdos1960evolution}. Equivalently, at fixed $L$, chaotic behavior sets in once $\tilde{M}$ exceeds $\mathcal{O}(1/L)$. This structural crossover in the $(\tilde{M}, L)$ plane underlies the system-size-dependent onset of chaos observed in our study at small $\tilde{M}$. 

At $\tilde{M}=1$ and $h=0$, the Hamiltonian in Eq.~\ref{Hamiltonian} can be rewritten as 
\begin{equation}
H=\frac{1}{2M}S_z^2 + \frac{1}{L}S_x,
\end{equation}
up to an additive constant, where $S_x=\sum_i X_i$ and $S_{z}=\sum_i Z_i$. The Hamiltonian is symmetric under the permutation group $S_L$ (relabeling spin indices) and commutes with the total spin squared $S^2 \equiv S_x^2 + S_y^2 + S_z^2$. As a consequence, the vector space can therefore be decomposed into
\begin{equation}
V=\sum_{S={S_{\rm min}}}^{L/2} V_S \otimes V_P,
\end{equation}
where $S_{\rm min}=0(1/2)$ if $L$ is even (odd). $V_S$ forms a spin $S$ irrep of $SU(2)$ and $V_P$ is some irrep of $S_L$ representing the multiplicity of the $S$ representation. On this decomposition, $H$ acts as $\sum_{S}^\oplus H_S\otimes \mathbb{I}_{V_P}$. To explicitly see this, we can use the following three-spin/qubit example. The space decomposes into irreps of $SU(2)$ as $\frac{3}{2}\oplus \frac{1}{2}\oplus \frac{1}{2}$. In the following basis
\begin{eqnarray}
&\vert \frac{3}{2},m-\frac{3}{2}\rangle \propto S_+^m \vert 000\rangle;\, m=0,1,\dots 3\nonumber\\
&\vert \frac{1}{2},-\frac{1}{2}, \alpha=0\rangle \propto (|01\rangle- |10\rangle)|0\rangle \nonumber\\&\vert \frac{1}{2},+\frac{1}{2}, \alpha=0\rangle \propto (|01\rangle- |10\rangle)|1\rangle\nonumber\\
&|\frac{1}{2},\frac{1}{2},\alpha=1\rangle
\propto
\sqrt{\frac{2}{3}}\,|110\rangle
-\frac{1}{\sqrt{6}}(|101\rangle + |011\rangle) \nonumber\\
&|\frac{1}{2},-\frac{1}{2},\alpha=1\rangle
\propto
\frac{1}{\sqrt{6}}(|100\rangle + |010\rangle)
-\sqrt{\frac{2}{3}}\,|001\rangle\nonumber
\end{eqnarray}
where the $\alpha$ index represents the multiplicity sector, $S_+$ is the collective spin raising operator, Hamiltonian has the form 
\begin{equation}
H=\left[\begin{array}{ccc}
H_{\frac{3}{2}} & 0 & 0\\
0 & H_{\frac{1}{2}} & 0\\
0 & 0 & H_{\frac{1}{2}}
\end{array}\right]\equiv H_\frac{3}{2}\oplus H_{\frac{1}{2}}\otimes \mathbb{I}_2.
\end{equation}
In a similar way, the $L=4$ case decomposes as $H_{2}\oplus H_{0}\otimes\mathbb{I}_{2}\oplus H_{1}\otimes\mathbb{I}_{3}$.

The permutation symmetry and the commutation with $S^2$ lead to extensive Hilbert space fragmentation with maximal fragment size upper bound by $L+1$ (corresponding to the $S=\frac{L}{2}$ sector). Less important here, the permutation symmetry guarantees extensive degeneracy from the multiplicity space acting as a zero mode.
The small fragments imply that the initial states undergo very restricted expansion in the Hilbert space, resulting in the absence of thermalization. As the bonds are removed when $\tilde{M}<1$, the symmetries are lost, resulting in weak coupling of the sectors. Eigenstates still have large weights in one block, resulting in an eigenstate ordering inherited from the nearby $\tilde{M}=1$ point. 

When $\tilde{M} = 1 - \epsilon$ for small $\epsilon > 0$, i.e., just below the maximally connected limit, the Hamiltonian can be viewed as a perturbation of the LMG model by a term
\begin{equation}
    V = \frac{2J}{L(L-1)\tilde{M}} \sum_{\langle ij \rangle_{\rm del}} Z_i Z_j,
\end{equation}
where the sum runs over the $\epsilon L(L-1)/2$ deleted bonds. Before the perturbation, the LMG energy eigenvalues depend only on the $S^2$ eigenvalue, $E = E(s)$. The perturbation introduces off-diagonal blocks coupling different $s$ sectors. For $s$ away from the fully polarized limits, each $s$ sector is exponentially degenerate, with multiplicity
\begin{equation}
    m(s) = \frac{2s+1}{L/2 + s + 1} \binom{L}{L/2 - s}.
\end{equation}
Treating the degenerate eigenstates as random vectors in the computational basis with typical coefficient magnitude $\sim 2^{-L/2}$, the off-diagonal blocks coupling sectors $s$ and $s'$ have dimensions $m(s) \times m(s')$ (exponentially large in $L$) and matrix elements of order $\langle s | V | s' \rangle \sim \mathcal{O}(\sqrt{\epsilon}/2^{L/2})$. Although individual matrix elements are exponentially small, the coupling between exponentially large degenerate blocks leads to a lifting of the block-diagonal structure and mixing of eigenstates at large $L$. While this constitutes only a plausibility argument for the emergence of chaos at any finite $\epsilon$, our numerical results are consistent with this picture.

While an atypical initial state, such as a direct product state, will have overlap with an extensive number of energy eigenstates, taking it to an infinite temperature diagonal ensemble at equilibration, the eigenstate ordering results in overlaps with only a restricted set of eigenstates, resulting in the diagonal ensemble deviating from the infinite temperature ensemble. We explicitly show this distinction through the inverse participation ratio (IPR) of such product state in the basis of eigenstates (Fig.~\ref{ipr_eigesnstate} in Appendix~\ref{ipr_appendix}).

In the rest of this section, we demonstrate that chaotic dynamics in such a Hamiltonian can affect QAOA performance when this Hamiltonian is employed as an additional term in QAOA protocol.

\subsection{QAOA with random Ising driver\label{QAOA}}
\begin{figure}[htbp]
		\includegraphics[width=\columnwidth]{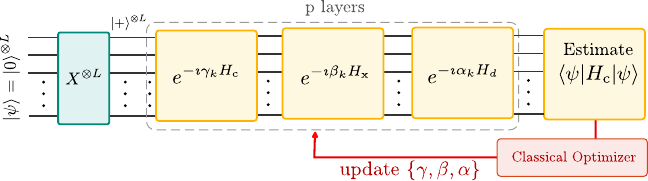}\caption{\label{qaoa_schematic}An outline of QAOA protocol. The three-term variational circuit is parameterized by $\{\alpha_k,\beta_k,\gamma_k\}$. The entire circuit consists of $p$ such layers. The input state of the ansatz is taken as $|+\rangle^{\otimes L}$. The energy of the cost function is optimized with a classical optimizer and the parameters are iteratively updated using gradient descent. }
\end{figure}

Here, we empirically demonstrate, in the specific setting of QAOA, that chaos in the instantaneous Hamiltonian can be relevant to the circuit performance. We consider the problem of minimizing a classical Ising cost Hamiltonian of the form
\begin{align}
    \label{qaoa_problem}
    H_{\rm c} &= \sum_{i,j  \in C} W_{ij} Z_iZ_j,
\end{align}
defined on a random graph $C$ with $L$ nodes. Several difficult classical optimization problems~\cite{barahona1982computational}, such as MAXCUT and SAT, can be cast as ground-state search of such an Ising Hamiltonian~\cite{lucas2014ising}. For the numerical demonstrations in this section, we sample the graph $C$ from a set of ER graphs and sample the edge weights $W_{ij}$ uniformly from $\{+1,-1\}$.  The standard QAOA ansatz alternates between the cost unitary $e^{\imath \gamma_k H_{\rm c}}$ and a mixer unitary $e^{\imath \beta_k H_{\rm x}}$ where $\gamma_k , \beta_k$ are variational parameters associated with the $k$-th layer which are to be optimized. The entire circuit consists of $p$ such layers. Here, we modify this QAOA protocol by adding an additional driver unitary $e^{\imath \alpha_k H_{\rm d}}$ in each layer as illustrated in Fig.~\ref{qaoa_schematic}. 
$H_{\rm d}$ is a mixed-field Ising model on an ER graph with Pauli-$YY$ interaction given as,
\begin{align}
\label{qaoa_Hd}
    H_{\rm d}^{YY} &= -\frac{JL}{M} \sum_{i,j}A_{ij} Y_iY_j -g\sum_i X_i. 
\end{align}
Its spectral properties are identical to those of the annealing Hamiltonian (Eq.~\ref{Hamiltonian} with $h=0$ for simplicity), and its chaotic character is therefore tunable through $\tilde{M}$.
Note that the adjacency matrices $C$ (of $H_{\rm c}$) and $A$ (of $H_{\rm d}$) are chosen independently. 
We take the initial state for the ansatz as $|+\rangle^{\otimes L}$ and set $J=g=1$. The values of $\alpha,\beta, \gamma$ are optimized such that $\langle H_{\rm c} \rangle $ is minimized. These unitaries need to be implemented on quantum devices using a Trotterized approach~\cite {elben2023randomized,Chen_2024}. 

\begin{figure}[htbp]
		\includegraphics[width=\columnwidth]{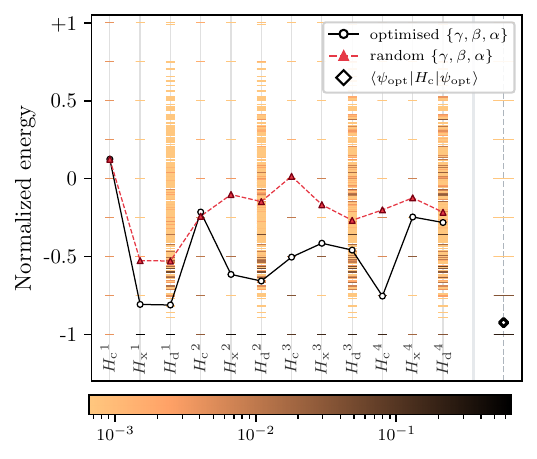}\caption{\label{trajectory}Trajectory of energy expectation value of instantaneous state in QAOA dynamics through the eigen-spectrum of instantaneous QAOA Hamiltonian. The thickness of dashes is proportional to the degeneracy of the eigenstates. The ansatz is a four-layer quantum circuit. The label at the bottom of each stack shows the instantaneous Hamiltonian and the corresponding layer (in superscript). The diamond marker in the right-most stack marks the final energy expectation value of the optimized state, i.e., $\langle \psi_{\rm{opt}}|H_{\rm c}|\psi_{\rm{opt}} \rangle$. The color indicates the overlap between the optimized state and the instantaneous eigenspectrum. $ L=8$. The problem graph has a connectivity $\tilde{M}_{\rm c}=0.4$. The driver (Eq.~\ref{qaoa_Hd}) has a connectivity $\tilde{M}=0.6$.}
\end{figure}

To understand the state evolution in such a circuit, we trace the trajectory of the energy expectation value of the instantaneous state at each step of a four-layer QAOA ansatz (Fig.~\ref{trajectory}). The background horizontal lines show the instantaneous eigenspectrum, colored based on the overlap with the optimized state $|\psi_{\rm opt}\rangle$ ; the darkest lines, concentrated at the low-energy end of $H_c$, indicate that $|\psi_{\rm opt}\rangle$ is strongly supported on low-energy states. For optimized parameters (black line), the trajectory passes through the bulk of the spectrum, with significant overlap onto mid-spectrum eigenstates of $H_{ d}$. This is qualitatively distinct from adiabatic evolution, in which the state remains confined to the low-energy manifold at all times. Nevertheless, the optimized circuit successfully drives the final state to near-ground-state energy, as indicated by the diamond marker. For randomly chosen parameters (red dashed line), the energy expectation value shows no systematic drift toward low energy, fluctuating throughout the bulk of the instantaneous spectrum.

\begin{figure}[htbp]
		\includegraphics[width=\columnwidth]{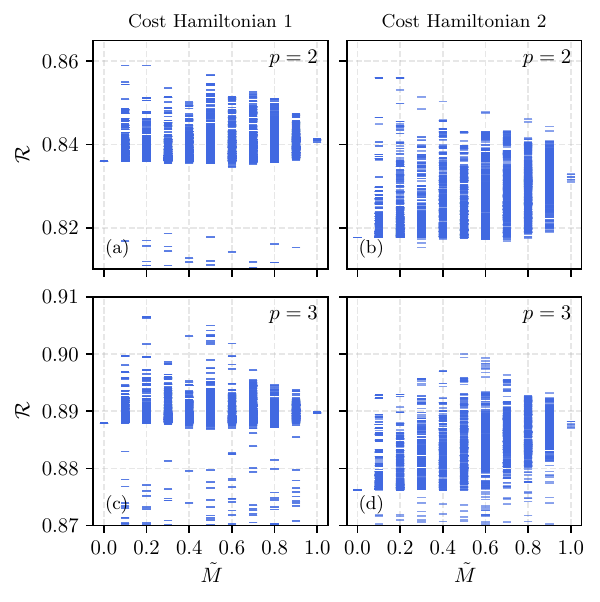}\caption{\label{qaoa_scatter} Approximation ratio of the optimized state for a QAOA circuit with $L=8,p=2$ (a,b) and $L=8,p=3$ (c,d). Keeping the problem $H_{\rm c}$ fixed, for each $\tilde{M}$, we show the QAOA optimized approximation ratio for $1000$ random choices of $H_{\rm d}$ at that $\tilde{M}$. The two columns describe data for two independently sampled problems $H_{\rm c}$ with $\tilde{M}_{\rm c}=0.4$.}
\end{figure}
We now assess the performance of this three-term QAOA protocol as a function of $~\tilde{M}$. We show the results for two randomly picked instances of the classical Ising Hamiltonian  $H_{\rm c}$ (Eq.~\ref{qaoa_problem}) on a random ER graph of $L=8$ sites with $\tilde{M}_{\rm c}=0.4$. Keeping $H_{\rm c}$ fixed, we sample $1000$ different realizations of $H_{\rm d}$ (Eq.~\ref{qaoa_Hd}) and optimize a circuit of $p=2,3$ layers. In each case, $600$ ADAM~\cite{kingma2015adam} gradient-descent steps were performed to minimize the cost function $E_{\rm{QAOA}}=\langle \psi | H_{\rm c }| \psi \rangle$ (energy expectation value of the final state with respect to the cost Hamiltonian). $E_{\rm{QAOA}}$ is evaluated numerically exactly and the gradient is obtained using finite difference method. To gauge the distribution of performance of the optimized QAOA, we compute the approximation ratio for each of the $1000$ instances as defined below,
\begin{equation}
\mathcal{R}=\frac{E_{\rm{max}}-E_{\rm{QAOA}}}{E_{\rm{max}}-E_{\rm{GS}}} \leq 1.
\end{equation}
Higher $\mathcal{R}$ implies better optimization performance. Here, $E_{\rm{GS}}$ and $E_{\rm{max}}$ are the ground state and maximum eigenvalue of $H_{\rm c}$ respectively. Figure~\ref{qaoa_scatter} shows the distribution of $\mathcal{R}$ values for the two cost Hamiltonians. Two features are particularly notable. First, $\mathcal{R}$ is higher at $\tilde{M}>0$ compared to the standard transverse-field mixer ($\tilde{M} = 0$) for the majority of driver instances, confirming that the additional Ising mixer improves QAOA performance. Further, the approximation ratio is larger at intermediate $\tilde{M}$ than at either extreme for most instances, where $H_{\rm{d}}$ exhibits chaotic spectral statistics. Note that the spread in $\mathcal{R}$ at $\tilde{M}=1$ is because of the variation in initial parameter value across multiple gradient-descent optimizations. This correlation suggests possible influence of the chaotic dynamics of $H_{\rm{d}}$ on QAOA performance, similar to what was observed for QAA~\cite{Schl_mer_2025} (see Appendix~\ref{qaoa_appendix} for additional numerical examples for system size up to $L=10$). 

The improvement in QAOA performance observed with  $H_{\rm d}$ containing $YY$ term can potentially be attributed to two distinct factors: the chaotic nature of $H_{\rm d}$, and the presence of terms that scatter between computational basis states with Hamming distance 2, enabling broader exploration of the solution space. To decouple these two contributions, we consider in Appendix~\ref{qaoa_appendix} an alternative driver Hamiltonian with Pauli-$ZZ$ coupling,
\begin{align}
\label{qaoa_Hd_z}
H_{\rm d}^{ZZ} &= -\frac{JL}{M} \sum_{i,j}A_{ij} Z_iZ_j - g\sum_i X_i,
\end{align}
in which the interaction term is diagonal in the computational basis but has identical spectral properties. We qualitatively obtain the same results, suggesting that the chaotic character of the entire $H_d$ as a whole, rather than the nature of its individual terms, leads to improved performance of the modified QAOA.

Having demonstrated a potential connection between circuit performance and chaos in the dynamics, in the subsequent sections, we present a thorough investigation of the dynamical properties of $H_{\rm{d}}^{ZZ}$ utilizing multiple complementary probes.

\section{Projected ensemble and emergence of deep thermalization \label{PE_section}}   

ETH predicts that the reduced density matrix of any small subsystem of an isolated quantum system evolving under a generic Hamiltonian relaxes to a Gibbs state determined solely by the initial energy density. This relaxation follows from the delocalization of initial local information across the many-body system, with scrambling as the underlying dynamical process. The subsystem's complement serves as an effective bath, whose degrees of freedom are traced out to yield the reduced density matrix. Under chaotic dynamics satisfying ETH, the resulting reduced state relaxes to a Gibbs ensemble, and all memory of the initial state is effectively erased from local observables. This framework is sufficient to predict expectation values of local subsystem observables. However, modern quantum simulator platforms can track all degrees of freedom simultaneously~\cite{bakr2009quantum,browaeys2020many,monroe2021programmable}, rendering the system-bath distinction artificial. Access to the bath state allows one to obtain pure states of the subsystem conditioned on the measurement outcome of the bath. The ensemble of such conditional pure states is the projected ensemble (PE), and constitutes a physically natural unraveling of the reduced density matrix. While a density matrix admits infinitely many unravelings, selection of a natural projective measurement basis is physically motivated. Unlike the density matrix, which encodes only the first moment of the state ensemble, the PE encodes the full distribution of quantum states in the Hilbert space, providing access to all higher moments and thereby to fluctuations beyond mean observable values. This motivates a distinction between thermalization at the level of local expectation values and the more fundamental notion of \textit{deep thermalization}, which concerns convergence of the underlying quantum state distribution in the projected ensemble to the Haar measure~\cite{lucas2023generalized,chang2025deep,mark2024maximum,chan2024projected,ippoliti2022solvable,bhore2023deep,manna2025projected,claeys2022emergent,yu2025mixed,liu2024deep}.

Consider a system of $L$ sites partitioned into two subsystems, $A$ and $B$, with $L=L_A+L_B$. Under unitary evolution, the entire system is described by a pure state, $|\psi_{AB} \rangle \in \mathcal{H}_A \otimes \mathcal{H}_B$ at all times. We consider initial states of the form
\begin{equation}
\label{projected_initstate}
		|\psi_{AB}(t=0)\rangle = \frac{1}{\sqrt{2^L}}\bigotimes_{i=0}^L ( |0 \rangle + \imath |1\rangle),
\end{equation}
evolved under the Hamiltonian in Eq.~\ref{Hamiltonian} (with $J=1,g=1$ and $h=0.1$). The state has $0$ energy expectation for this Hamiltonian. In the absence of any conserved charges (other than energy), the subsystem properties are expected to relax to those of a Gibbs ensemble $e^{-\frac{1}{T} H}$ where $T$ is the effective temperature of the initial state. As the Hamiltonian is traceless, $T=\infty$, and the subsystem properties of the time-evolved states (with Eq.~\ref{projected_initstate} as initial state) relax to that of the identity density matrix. 

We perform simultaneous projective measurements of Pauli-$Z$ on all qubits in subsystem $B$ on copies of $| \psi_{AB} \rangle$ and collect the corresponding pure state $| \psi_A(b_i) \rangle$ in $A$ conditioned on the measurement outcome $b_i$,
\begin{equation}
|\psi_{A,b_i}\rangle =\frac{1}{\sqrt{p(b_i)}}\Pi_{b_i} |\psi_{AB} \rangle =  | \psi_A(b_i) \rangle \otimes |b_i\rangle,
\end{equation}
where $\Pi_{b_i}=\mathbb{I}\otimes |b_i\rangle \langle b_i|$ and $p(b_i)=\langle \psi_{AB}|\Pi_{b_i}|\psi_{AB} \rangle$. The state $|\psi_{A,b_i}\rangle$ arises with probability $p(b_i)$ and the ensemble of these states, described by $\mathcal{E}_{\mathrm{PE}}=\{p(b_i),|\psi_{A,b_i}\rangle\}$, is the PE. When seen as an ensemble of the states in $A$, its distribution over the Hilbert space of $A$ can be characterized by its $k$-th moment as, $\rho^{(k)}_{\mathrm{PE}}=\sum_{b_i}p(b_i) (|\psi_A(b_i)\rangle \langle \psi_A(b_i)|)^{\otimes k}$. Given the PE, these can be computed using the $b_i$ labels on the ensemble states. The $k$th moment of any observable's expectation value can be obtained as $\langle \hat{O}^{\otimes k} \rangle_{\mathrm{PE}} = \mathrm{Tr}[\rho^{(k)}_{\mathrm{PE}} \hat{O}^{\otimes k}]$.

To characterize deep thermalization, one must identify the universal ensemble toward which the PE converges in the limit of large system sizes $L$ and long times. Under unitary evolution with $H$ from the initial state in Eq.~\ref{projected_initstate}, if $H$ is chaotic, the PE asymptotically (in $L,t$) approaches the Haar ensemble~\cite{cotler2023emergent,choi2023preparing,harrow2013church,zyczkowski2001induced}. The choice of the measurement basis does not affect PE convergence in this case~\cite{manna2025projected,mark2024maximum}.
In the presence of conservation laws or constraints, the asymptotic projected ensemble reached at $L,t \to \infty$ deviates from the Haar ensemble and is determined by the charges of the initial state~\cite{manna2025projected,chang2025deep,mark2024maximum} as well as any correlation between measurement outcomes and the conserved charges. The degree of deep thermalization is thus quantified by the deviation of the PE moments from those of the Haar ensemble.
   
\begin{figure}[htbp]
	\includegraphics[width=\columnwidth]{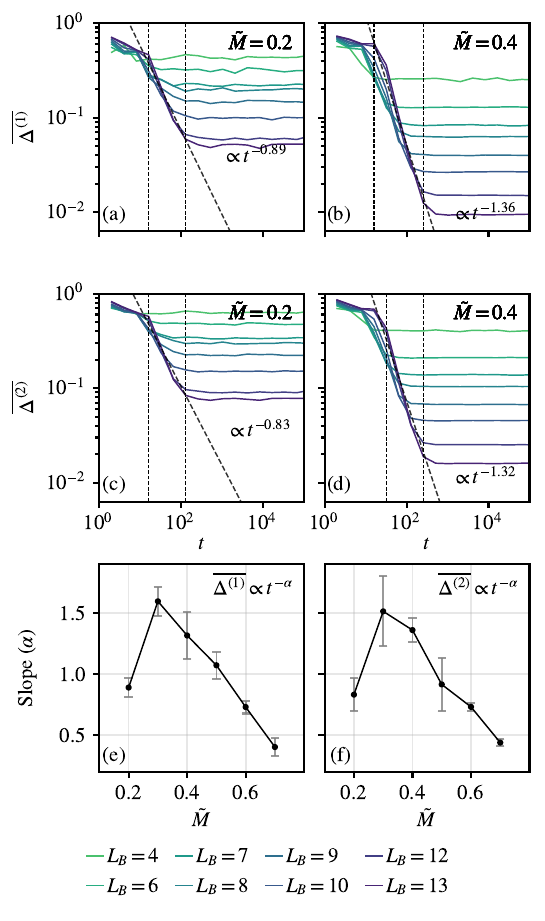}
	\caption{(a)-(b) Evolution of $\overline{\Delta^{(1)}}$ with time for $\tilde{M}=0.2$ and $0.4$, respectively.  (c)-(d) Evolution of $\overline{\Delta^{(2)}}$ with time for $\tilde{M}=0.2$ and $0.4$, respectively. All data represented here are averaged over $500$ graphs. Power-law fits for $L=14$ are shown with black dashed lines, and the corresponding exponents are noted in the plot. Vertical black dashed lines indicate the region where the fitting was done to obtain the power-law exponents. (e)-(f) Power-law decay exponent of the trace distance, obtained from a linear fit of $\log_{10} \Delta^{(k)}$ vs $\log_{10} t$, as a function of $\tilde{M}$ (error bars denote the standard error of the slope from the least-squares log-log fit.). $L_A = 2$; the Hamiltonian parameters in Eq.~\ref{Hamiltonian} are $g = 1.0$, $h = 0.1$. \label{timeevo_trace_dist}}
\end{figure}

\begin{figure*}[htbp]
	\includegraphics[width=\textwidth]{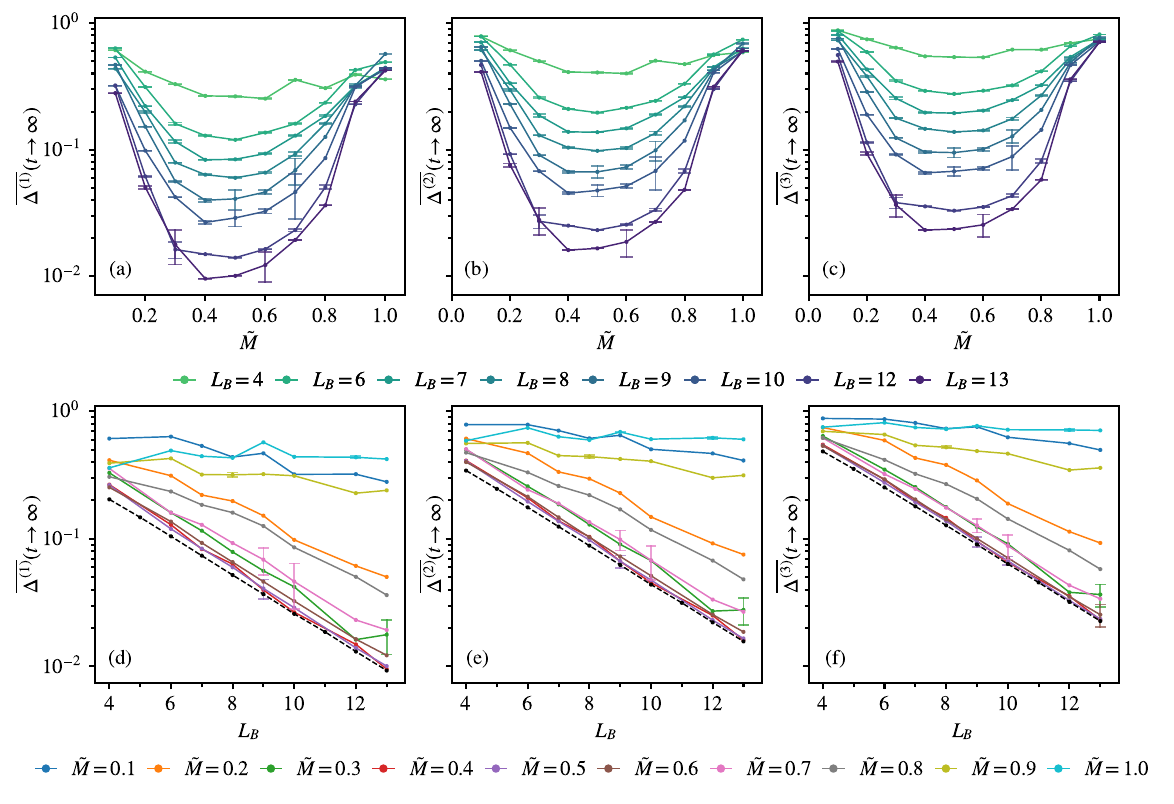}
	\caption{(a)-(c) Asymptotic ($t=10^9J^{-1}$) trace distance of first three moments of PE from Haar ensemble as a function of $L_B$. The asymptotic trace distance gets suppressed with increasing $L_B$ for all $\tilde{M}$. Error bars represent the standard error of the mean across realizations at a fixed $t$. (d)-(f) The same data plotted to show the the scaling of $\overline{\Delta^{(k)}}(t \to \infty)$ with $L$ across all $\tilde{M}$ for first three moments of PE. The error bars are not shown when they are comparable to the marker size. Each data point shown represents an average over 500 graph realizations, calculated for a subsystem of size $L_A=2$ with Hamiltonian parameters (Eq.~\ref{Hamiltonian}) $g=1.0$ and $h=0.1$. The black dashed line shows the trace distance of an empirical ensemble of $500$ random states (sampled from Haar random states of system size $L$) from the corresponding moments of the Haar ensemble. \label{asymptotic_connectance}}
\end{figure*}

To compare the Haar ensemble and the PE, we use the trace distance between corresponding moments. Given two ensembles, we define the trace distance between their $k$-th moments as,
\begin{equation}
	\label{tr_dist_eq}
\Delta^{(k)}(\mathcal{E},\mathcal{E'}) = \frac{1}{2}\mathrm{Tr}\sqrt{(\rho^{(k)}_{\mathcal{E}}-\rho^{(k)}_{\mathcal{E'}})(\rho^{(k)}_{\mathcal{E}}-\rho^{(k)}_{\mathcal{E'}})^\dagger}.
\end{equation} 
We will denote quantities averaged over multiple graph realizations with overbar. The results presented here are averages over $500$ graph realizations.  Figure~\ref{timeevo_trace_dist} (a-d) show the evolution of trace distances $\Delta^{(1)}(\mathcal{E},\mathcal{E}_{\rm{Haar}})$ and $\Delta^{(2)}(\mathcal{E},\mathcal{E}_{\rm{Haar}})$ with time for different $\tilde{M}$. The distances decay with time in all cases, ultimately saturating to a system-size-dependent asymptotic value, $\overline{\Delta^{(k)}}(t \to \infty)$. Figure~\ref{asymptotic_connectance} shows the dependence of the asymptotic distance on the system size. Both the initial decay with time and the $L$ dependence of the asymptotic distance show clear dependence on $\tilde {M}$.

\paragraph*{Convergence of PE to the Haar ensemble with time:} For intermediate $\tilde{M}$, the trace distance between the PE and the Haar ensemble decays as a power law in time, with an $\tilde{M}$-dependent exponent $\alpha$ (Fig.~\ref{timeevo_trace_dist}(a)--(d)). The convergence is fastest in the chaotic regime, with the largest exponent ($\alpha \approx 1.6$) occurring at $\tilde{M} \approx 0.3$ (Fig.~\ref{timeevo_trace_dist}(e)--(f)). For comparison, the nearest-neighbor mixed-field Ising model at a specific chaotic parameter point (where it is expected to be chaotic) exhibits an exponent of $\approx 1.2$~\cite{cotler2023emergent}. The exponent decreases on either side of $\tilde{M} \approx 0.3$, toward both the sparse and dense limits. The power-law exponents extracted from the first and second moments of the PE agree within error bars, indicating that the exponent governing the approach to thermalization is independent of the moment order, a feature previously reported in Refs. ~\cite {cotler2023emergent, lucas2023generalized}. At extremal values ($\tilde{M} \leq 0.1$, $\tilde{M} \geq 0.8$), the trace distance exhibits a slower approach to its asymptotic value, accompanied by pronounced plateaus, suggesting that thermalization proceeds via distinct dynamical timescales in these regimes.

\paragraph*{Asymptotic (late-time) trace distance, $\overline{\Delta^{(k)}}(t \to \infty)$:} We compare the late-time asymptotic trace distances $\Delta^{(k)}$ for $k=1,2,3$ in Fig.~\ref{asymptotic_connectance}(a)-(c) as a function of $\tilde{M}$ for multiple $L$. Figure~\ref{asymptotic_connectance}(d)-(f) show the $L$ dependence of $\overline{\Delta^{(k)}}(t \to \infty)$. 

At fixed $L$, the asymptotic trace distance is largest at extremal $\tilde{M}$ and decreases toward intermediate connectance, reaching a minimum at $\tilde{M} \approx 0.4$. At $\tilde{M} = 0.4$--$0.5$, the trace distance approaches the same obtained for Haar random states of the same system size (black dashed lines in Fig.~\ref{asymptotic_connectance}(d)--(f)). The larger trace distance at $\tilde{M}$ away from $0.4$ suggests that these Hamiltonians possess approximate conservation laws that impose structure on the eigenstates, preventing full ergodicity (as discussed in Sec.~\ref{QAA_section}). We find no evidence of slow time-dependent drift in the asymptotic region (Fig.~\ref{timeevo_trace_dist}(a)--(d)), confirming that the trace distance has genuinely converged rather than remaining in a slow transient. On the sparse side ($\tilde{M} \ll 0.4$), disconnected clusters give rise to local conserved quantities, accounting for some of the observed deviation from chaos. On the dense side ($\tilde{M} \gtrsim 0.4$), the approximate permutation symmetries inherited from the near-LMG structure provides the relevant approximate conservation laws that impede PE convergence.

The asymptotic trace distance $\overline{\Delta^{(k)}}(t \to \infty)$ decays exponentially with $L$ at intermediate $\tilde{M}$ over the accessible system sizes, suggesting convergence of the PE toward the Haar ensemble in the thermodynamic limit and deep thermalization in chaotic models. In contrast, at $\tilde{M} \to 0, 1$, the decay with $L$ is significantly slower, while the fastest decay occurs at intermediate $\tilde{M}$. This qualitative behavior is consistent across the first, second, and third moments of the PE. The fact that the trace distance decays with $L$ even at $\tilde{M} = 0.1$ and $0.9$, albeit slowly, is tentative evidence that the system may be chaotic throughout, except at the integrable limits $\tilde{M} = 0, 1$, a conclusion consistent with Ref.~\cite{PhysRevResearch.7.013146}.

So far in this section, we have investigated the PE's approach to the Haar ensemble for time evolutions starting from a specific zero-energy expectation-value initial state (Eq.~\ref{projected_initstate}). At $\tilde{M}=1.0$, the Hamiltonian has a permutation symmetry which is preserved by this state, as a result of which the instantaneous state has finite overlaps with only a small number of energy eigenstates of the $\tilde{M}=1$ Hamiltonian. However, the results shown in Fig.~\ref{asymptotic_connectance} stay valid even if we consider other zero energy expectation value initial states that lack permutation symmetry. In Appendix~\ref{sec:PE_random}, we show analogous results for initial direct product states with limited or no permutation symmetry, demonstrating the robustness of the results presented in Fig.~\ref{asymptotic_connectance}.

In summary, we find that finite-system studies of the PE show a clearly faster onset of chaos (through convergence of the PE to the Haar ensemble) with increasing system size at intermediate connectivity $\tilde{M}\sim 0.4-0.6$. Away from this, the approach of PE to the Haar ensemble is slower, likely because of approximate conserved charges or symmetries. In the latter cases, in accessible systems, the asymptotic PE approaches the Haar ensemble as system size increases, indicating that the number of approximately conserved charges and symmetries becomes subextensive. Measurement of the projected ensemble and its higher moments has been demonstrated in quantum simulators for 1D lattice systems with up to $25$ qubits~\cite{cotler2023emergent}. Extending such protocols to random-graph connectivity presents additional experimental challenges, particularly in implementing the requisite couplings on hardware with limited connectivity and in the measurement-sampling overhead needed to resolve higher moments of the projected ensemble. Nevertheless, our finite-size results suggest that the trace distance of the PE from the Haar ensemble provides a clear and resolvable diagnostic of the chaotic crossover, capturing the $\tilde{M}$-dependent convergence rate and the moment-independence of the thermalization exponent. These signatures are visible at system sizes accessible to near-term quantum devices, making this a tractable target for experimental investigation.

\section{Partial Spectral Form Factor\label{pSFF_section}}
Having characterized the emergence of chaos through the properties of the projected ensemble, we now turn to a complementary diagnostic: the correlations within the energy spectrum itself. To this end, we investigate the partial spectral form factor (pSFF), which generalizes the spectral form factor (SFF) to subsystems and serves as a probe of spectral correlation, a key signature of quantum chaos. Unlike the SFF, the pSFF is more readily accessible in experiments. In this section, we utilize the pSFF to study the localization-to-chaos crossover as a function of connectivity ($\tilde{M}$) and explore its dependence on both system and subsystem size in different connectivity regimes.

    Before discussing the pSFF, we briefly review the SFF~\cite{PhysRevX.8.021062,Abhishodh_2019,vsuntajs2020quantum,PhysRevLett.121.264101}. The SFF is a spectral diagnostic of quantum chaos that encapsulates both short- and long-range spectral correlations, providing information beyond those contained in level-spacing statistics (which captures only adjacent level spacings). The SFF is the Fourier transform of the eigenvalue density-density correlation, defined as
    \begin{align}
    \label{SFF}
        K(t)=\frac{1}{d^2}\sum_{i,j=1}^{d}e^{i(E_i-E_j)t} =\frac{1}{d^2} \mathrm{Tr}[\mathcal{T}(t)] \mathrm{Tr}[\mathcal{T}^{\dagger}(t)]\ ,
    \end{align}
    where $d$ is the Hilbert space dimension of the full system and $E_i$ is $i$-th eigenvalue of the Hamiltonian. $t$ is the Fourier variable conjugate to energy, with units of time. The time evolution unitary operator is represented as, $\mathcal{T}(t)=e^{-\imath H t}$. The prefactor ensures $K(0)=1$. In a chaotic spectrum (obeying WD statistics), the SFF  decays from $1$ at $t=0$ until the Thouless time $t_{Th}\approx O(\delta E ^{-1})$ where $\delta E$ is the mean level spacing. This results in the SFF dropping to below $1/d$ at intermediate time. This is followed by a linear ramp that saturates at $1/d$ at the plateau time (the time at which the plateau in SFF begins), $t_p\approx O(2^L)$. The dip in SFF (referred to as a correlation hole) is caused by level repulsion and spectral rigidity in chaotic systems as predicted by random matrix theory (RMT). This ramp-and-plateau structure is a signature of the chaotic spectrum. For a spectrum following Poisson statistics, the SFF is expected to gradually reach its asymptotic value after an initial transient period without any correlation hole or ramp. For chaotic systems, the asymptotic SFF is given as $K(t \to \infty)=1/d$ in the absence of any degeneracy in the eigenspectrum. Recently, SFF has been experimentally measured in a 5-qubit superconducting device using a randomized measurement protocol~\cite{dong2025measuring}. Reference ~\cite{PhysRevResearch.7.013146} analyzed SFF in the model studied here for $h=0$. In the intermediate regime of both $\tilde{M}$ and $g$, the SFF converges with increasing $L$ to the characteristic dip-ramp-plateau structure predicted by RMT, indicating chaotic spectra. Conversely, chaos is suppressed at the extreme limits of connectivity. In both the sparsely connected ($\tilde{M} \to 0$) and densely connected ($\tilde{M} \to 1$) regimes, the development of the correlation hole is slower with increasing $L$. Similarly, the correlation hole vanishes for both very small and very large $g$, indicating a breakdown of RMT statistics. 
        
    Despite recent experimental demonstrations of SFF measurement on superconducting qubit arrays, a significant practical challenge remains regarding the requirement of global control and site-resolved operations across the entire quantum system. In practice, implementing such fully controlled rotations on all qubits becomes resource-intensive as system size increases.  The partial spectral form factor (pSFF) was recently introduced as a scalable alternative~\cite{joshi2022probing}, which, unlike the full SFF, requires controlled operations that scale only within a smaller subsystem~\cite{joshi2022probing,dong2025measuring}. The pSFF is described in terms of the time-evolution operator restricted to a subsystem. The pSFF can be defined as
 	\begin{align}
    \label{pSFF_defn_eq}
            K_A(t) &= \frac{1}{d d_A} \sum_{i,j} e^{\imath (E_i - E_j)t} \mathrm{Tr}_B[\rho_B(E_i)\rho_B(E_j)] \nonumber \\
            &= \frac{1}{d d_A} \mathrm{Tr}_B [\mathrm{Tr}_A[\mathcal{T}(t)] \, \mathrm{Tr}_A[\mathcal{T}^{\dagger}(t)]],
        \end{align}
 where $\rho_B(E_i) = \mathrm{Tr}_A \left[ |E_i\rangle \langle E_i| \right]$ is the reduced density matrix of subsystem $B$ for energy eigenstate $|E_i\rangle$. pSFF captures the correlation of both eigenstates and eigenvalues. The normalization factor in Eq.~\ref{pSFF_defn_eq} ensures $K_A(0)=1$.

  \begin{figure}[htbp]
 			\includegraphics[width=\columnwidth]{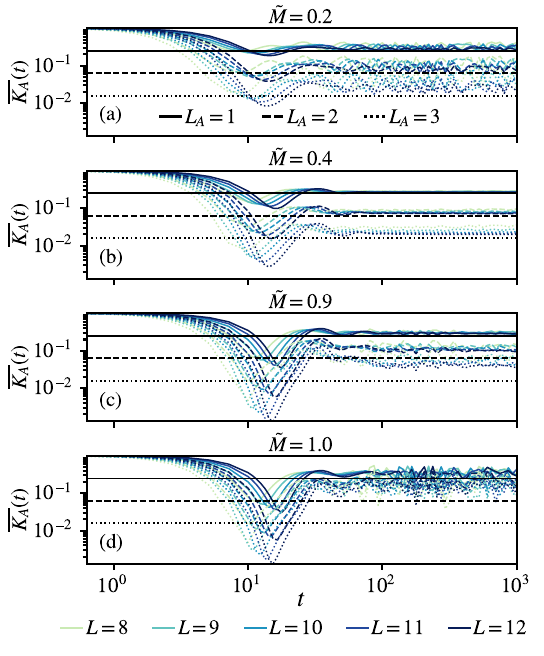}
 			\caption{Time evolution of the pSFF, $\overline{K_A}(t)$, for various system sizes ($L$, different colors) and subsystem sizes ($L_A$, different line styles). Panels show different connectivities: (a) $\tilde{M}=0.2$, (b) $\tilde{M}=0.4$, (c) $\tilde{M}=0.9$, and (d) $\tilde{M}=1.0$. The solid, dashed, and dotted lines correspond to $L_A=1, 2,$ and $3$, respectively. The horizontal black lines denote the expected asymptotic values for a chaotic system, $1/d_A^2$. All data are averaged over $500$ graph realizations (denoted by an overbar) with Hamiltonian parameters (Eq.~\ref{Hamiltonian}) of $g=1.0$ and $h=0.1$.\label{psFF_avg_graphs}}
 \end{figure}

     We now investigate the pSFF for the Hamiltonian in Eq.~\ref{Hamiltonian}, setting $g=1$ and $h=0.1$ with the goal of understanding the manifestation of localization, integrability and chaos as well as the finite-size effect in these cases. 
The SFF can be computed directly from the eigenvalue spectrum, allowing spectral unfolding on mid-spectrum eigenvalues to remove non-universal level-density variations. However, in an experimental setting, the full eigenspectrum is usually inaccessible, precluding spectral unfolding. In the following, we therefore compute the pSFF for the full spectrum without unfolding.
In spite of this, we find that the pSFF reveals clear chaotic behavior at intermediate connectivity and distinct deviations at extremal connectivity, making the pSFF another route to investigate the model in scalable experimental systems. 

We analyze the behavior of pSFF at multiple values of $\tilde{M}$ with the results presented in Fig.~\ref{psFF_avg_graphs}. The pSFF is averaged over $500$ ER graph realizations. For chaotic systems, RMT predicts~\cite{joshi2022probing} (in the absence of degeneracy) $K_A(t \to \infty)\approx K(t \to \infty )+1/d_A^2 = 1/d + 1/d_A^2$. For $L_A \ll L$, $K_A(t\to \infty) \approx 1/d_A^2$ i.e., the asymptotic (late-time) pSFF shows a subsystem size ($L_A$) dependent shift. In Fig.~\ref{psFF_avg_graphs}, these asymptotic values (for chaotic spectra) are shown using black horizontal lines.

\paragraph*{pSFF at small $\tilde{M}$ :} For $\tilde{M} =0.2$, Fig.~\ref{psFF_avg_graphs}(a)  shows that there is no distinct correlation hole, and the pSFF reaches its asymptotic value after a transient period. The absence of a correlation hole indicates the lack of correlation in the eigenspectrum (consistent with Poisson statistics). There is also a persistent oscillation present at late-times, even after averaging over graph realizations. Apart from the late-time fluctuation in the pSFF, the mean asymptotic value is also larger than the expectation for a chaotic Hamiltonian. This can be attributed to quasi-degeneracies in the eigenspectrum (when adjacent level spacing falls below $\delta E$) \cite{PhysRevResearch.7.013146}. For $\tilde{M}$ close to 0, there are large fluctuation at all times because of the existence of disconnected clusters, which obscures the possible existence of any correlation hole.

\paragraph*{pSFF at intermediate $\tilde{M}$ :}  The pSFF shows qualitatively different behavior in the chaotic regime. In Fig.~\ref{psFF_avg_graphs}(b), the pSFF shows a dip (correlation hole) and a subsequent ramp-and-plateau structure. This correlation hole arises from energy-spectrum correlations, as expected in chaotic systems. The depth of the correlation hole increases with both $L$ and $L_A$. The time of the appearance of the correlation hole, $t_{\rm Th}$, increases with $L$ for fixed $L_A$. For $\tilde{M} =0.4$, Fig.~\ref{psFF_avg_graphs}(b) shows that the asymptotic $K_A$ shows very little fluctuation at late-time and approaches $\approx 1/d_A^2$. Additionally, there is no significant fluctuation in the pSFF at $t > t_p$ when no average over graph realizations is taken.
 
 \paragraph*{pSFF at large $\tilde{M}$ : } In the densely connected limit approaching all-to-all connectivity, i.e., $\tilde{M} = 0.9$,  Fig.~\ref{psFF_avg_graphs}(c) shows a dip in pSFF. In this case, the asymptotic pSFF shows small oscillation around a plateau value larger than $1/d_A^2$. This deviation from $1/d_A^2$ is caused by approximate degeneracy in the energy spectrum close to $\tilde{M}=1$.  

\paragraph*{pSFF at $\tilde{M}=1.0$ : } At $\tilde{M} = 1.0$, i.e., where the model is integrable, we observe a dip in the pSFF at an intermediate time, a feature that persists even in the $h=0$ limit. The pSFF eventually saturates to a value exceeding the RMT prediction for chaotic systems, suggesting a deviation from WD spectra. 
 
 \begin{figure}
 	\includegraphics[width=\columnwidth]{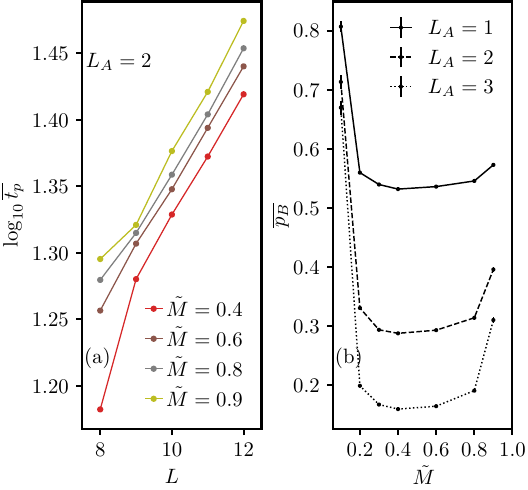}
 	\caption{Quantitative analysis of the pSFF. (a) Scaling of the plateau time, $\log_{10} t_p$, as a function of system size $L$ for several connectance, $\tilde{M}$, with a fixed subsystem size of $L_A=2$. (b) The late-time subsystem purity, $\overline{p_B}$, as a function of $\tilde{M}$ for a fixed system size $L=12$ and varying subsystem sizes, $L_A$. All data are averaged over $500$ graph realizations (denoted by an overbar) with Hamiltonian parameters (Eq.~\ref{Hamiltonian}) of $g=1.0$ and $h=0.1$. Here overbar denotes annealed average. \label{t_p_pSFF}}
 \end{figure}

\paragraph*{Scaling of the plateau time :}To further quantify the ramp-and-plateau behavior and its dependence on $L$ in the chaotic regime, we study the scaling of the plateau time, $t_p$ with $L$ for $L_A=2$. This time is determined numerically as the intersection of two linear fits: one to the logarithmic plot of the ramp ($t_{\rm Th}<t<t_p$) and another to the late-time plateau ($t>t_p$). For the system sizes accessible ($L = 8-12$), $t_p$ appears to scale exponentially with $L$, as shown in Fig.~\ref{t_p_pSFF}(a), with a slope that does not strongly depend on $\tilde{M}$ within the chaotic regime. The exponential scaling of the plateau time is a feature~\cite{joshi2022probing,dong2025measuring} of the SFF that is inherited by the pSFF. Our data in Fig.~\ref{t_p_pSFF}(a) is consistent with this behavior.

 \begin{figure}
 	\includegraphics[width=\columnwidth]{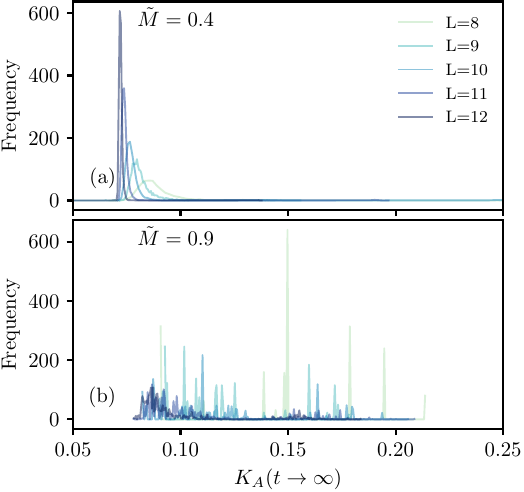}
 	\caption{Normalized distribution of the asymptotic (late-time) pSFF, $K_A(t \to \infty)$, computed over all possible two-site subsystems for different system sizes, $L$. Panel (a) shows the distribution in the chaotic regime, $\tilde{M}=0.4$. Panel (b) shows the distribution in the near-integrable regime, $\tilde{M}=0.9$. For all calculations, the subsystem size is fixed at $L_A=2$ and the Hamiltonian (Eq.~\ref{Hamiltonian}) parameters are $g=1.0$ and $h=0.1$. The data is presented here for $200$ graph realizations.\label{dist_pSFF_LA_2}}
 \end{figure}
 
\paragraph*{Entropy from pSFF :}Another quantity related to pSFF is the purity of the subsystem averaged over eigenstates, defined as $p_B=\mathbb{E}_{i}[\mathrm{Tr}(\rho_B(i) ^2)]$ where $\rho_B(i)$ is the reduced density matrix of the subsystem $B$ in the $i$-th eigenstate. This quantity captures the entanglement of subsystem with its complement and is related to the annealed subsystem Renyi entropy as $R_2=-\ln(p_B)$. The purity can be obtained from pSFF at $t \to \infty$ as $p_B=\lim _{t_p/t \to 0 }K_A(t)d_A$. Figure~\ref{t_p_pSFF}(b) shows $p_B$ as a function of $\tilde{M}$ for multiple $L_A$ and $L=12$. The purity is minimum at the intermediate $\tilde{M}$ i.e., subsystem $B$ is more mixed (more entanglement with $A$) in the chaotic regime compared to the integrable or localized regime. At both $\tilde{M} \to 0,1$, the purity goes up. The higher purity at $\tilde{M} \to 0,1$ is consistent with the expectation of lower entanglement in the localized or integrable eigenstates compared to the chaotic eigenstates. The purity of $B$ decreases (i.e., subsystem entanglement increases) with increasing $L_A$.

\paragraph*{ Fluctuation in the pSFF arising from the choice of subsystem : }To complete our analysis of the the pSFF,  we investigate the fluctuation in the pSFF arising from the choice of subsystem. We compute pSFF for $L_A=2$ subsystems over $200$ graph realizations for multiple $L$. In each realization, we choose all possible subsystems consisting of two sites as $A$ and the rest of the sites constitute $B$. Figure~\ref{dist_pSFF_LA_2}(a) shows the distribution of asymptotic pSFF values for $L_A=2$ and multiple $L$ at $\tilde{M}=0.6$. The variance reduces with increasing $L$. At large $L$, the choice of sites constituting the subsytem does not affect the asymptotic pSFF in the chaotic regime. The peak also shows a shift toward the expected value (in chaotic spectra) $ \approx 1/d_A^2$ ($0.0625$ for $L_A=2$) with increasing $L$. This behavior contrasts sharply with the near-integrable regime ($\tilde{M}=0.9$). Here, Fig.~\ref{dist_pSFF_LA_2}(b) shows a broad distribution for small $L$. The distribution becomes narrower with $L$ for fixed $L_A$. However, it is significantly broader compared to the chaotic case for the same $L$ and $L_A$.

\paragraph*{ Experimental implications :} Qualitative features of pSFF across dynamical regimes resemble those of SFF. As the pSFF can be experimentally obtained using randomized measurement (with lower resource requirements compared to SFF), we briefly comment on the number of measurements required to estimate pSFF with sufficient accuracy to capture the physics described in this section. As established in Ref.~\cite{joshi2022probing},  the relative error, $\epsilon$, of the pSFF measurement scales as the number of experimental shots, $N_m$, as $\epsilon \propto \sigma_A/{K_A \sqrt{N_m}}$ at each instant of time, where $\sigma_A$ is the variance in the shadow estimator of pSFF. This implies that the number of measurements needed to reach a desired precision, $N_m \propto \sigma_A^2/K_A^2$. The number of required measurement is inversely proportional to the instantaneous value of pSFF. This makes the estimation of the dip in the pSFF (i.e., around $t=t_{\rm Th}$) challenging for a large subsystem in a chaotic regime as $K_A(t \approx t_{\rm Th})  = O(10^{-3})$ in our model. Compared with this, the asymptotic value of pSFF serves as a better distinguishing factor, as the magnitude is typically one order of magnitude larger here ($t>t_p$). However, reaching the plateau requires long time evolution, making it susceptible to decoherence. The correlation hole, appearing at an earlier time, thus offers a more practical target despite its higher measurement overhead.

The fluctuation of the late-time pSFF over time, within a single experimental realization, provides a qualitative diagnostic that distinguishes chaotic from integrable Hamiltonian dynamics. The markedly lower variance of the asymptotic pSFF over subsystem choices in the chaotic regime further implies a reduced experimental overhead, requiring fewer independent realizations for accurate estimation of the pSFF relative to its integrable counterpart.
 
\section{Operator delocalization through Krylov complexity \label{Krylov_section}}
In a closed quantum system, generic Hamiltonian evolution causes initial operators to become increasingly complex over time. While probes like out-of-time-ordered correlators (OTOCs)~\cite{Maldacena_2016} can be studied for operator spreading, their reliance on a spatial separation metric is ill-suited for random graphs that lack an inherent local structure. A locality-independent quantification of complexity is the measure of the extent to which an operator spreads under time evolution in a basis composed of increasingly complicated operators~\cite{von2018operator,tan2025operator,parker2019universal}. In this section, we analyze the dynamics of operator spreading in the system through Krylov complexity. The time evolution of an operator is generated by applying the propagator constructed from the Liouvillian superoperator, and the resulting KC dynamics provide a diagnostic for quantifying chaos and localization.

First, we briefly review KC from the operator viewpoint (an alternative formulation based on state complexity has also been proposed \cite{balasubramanian2022quantum}). We start with an initial operator $\mathcal{O}(0)$. In Heisenberg picture, the time-evolved operator $\mathcal{O}(t)$ can be expressed as
\begin{align} 
\mathcal{O}(t)=e^{\imath Ht}\mathcal{O}(0)e^{-\imath Ht}=e^{\imath \mathcal{L}t}|\mathcal{O}(0)),
\end{align} 
where the Liouvillian $\mathcal{L}=(\mathbb{I} \otimes H - H^T \otimes \mathbb{I})$ and $|\mathcal{O})=\mathrm{vec}(\mathcal{O}) \in \mathcal{H}_d \otimes \mathcal{H}_d $ refers to the vectorized operator defined using
\begin{gather}
	\mathcal{O}=\sum_{i,j}\mathcal{O}_{i,j}|i\rangle \langle j| \nonumber \\
    \implies |\mathcal{O})=\text{vec} \left[\sum_{i,j}\mathcal{O}_{i,j}|i\rangle \langle j|\right]=\sum_{i,j}\mathcal{O}_{i,j}|i\rangle |j \rangle.
\end{gather} 
Here, $|i\rangle, |j \rangle$ refers to computational basis states. The action of $\mathcal{L}$ on a generic operator $\mathcal{O}$ is defined as $\mathcal{L}|\mathcal{O})=\mathrm{vec}([H,\mathcal{O}])$. Subsequently, we can define the Krylov subspace ($\mathcal{K}_\mathcal{O}$) of $\mathcal{L}$ associated with $\mathcal{O}$ as the minimal subspace to which $\mathcal{O}$ is restricted under time-evolution with $\mathcal{L}$ and is spanned by, 
\begin{align}
	\mathcal{K}_\mathcal{O}&=\text{span}\{\mathcal{L}^n|\mathcal{O})\}_{n=0}^{\infty}=\text{span}\{|\mathcal{O}),|\mathcal{L}\mathcal{O}),|\mathcal{L}^2\mathcal{O}),...\} \nonumber \\
	&=\text{span}\{\text{vec}(\mathcal{O}),\text{vec}([H,\mathcal{O}]),\text{vec}([H,[H,\mathcal{O}]])...\}.
\end{align}
The dimension of $\mathcal{K}_{\mathcal{O}}$ is denoted by $K$ and bounded~\cite{rabinovici2021operator} as $1 \leq K \leq d^2 - d + 1$. Evidently, the nested commutator operators in $\mathcal{K}_{\mathcal{O}}$ gets wider (have larger support) with increasing order. We define an inner product on operator space as $(A|B)=\mathrm{Tr}(A^{\dagger}B)$, and work with a complete orthonormal basis $\{|\mathcal{O}_n)\}_{n=0}^{K-1}$ in $\mathcal{K}_O$ obtained using Lanczos algorithm and implemented through successive full Gram-Schmidt orthogonalization  (see Appendix~\ref{sec:Lanczos}) starting from $|\mathcal{O}_0)=\mathrm{vec}(\mathcal{O}(0))$ i.e., the first element of the basis is the initial operator. This orthonormal basis in operator space is denoted as the Krylov basis. The Liouvillian takes a tri-diagonal form when expanded in $\{|\mathcal{O}_n)\}$, 
\begin{align}
	\mathcal{L} = \begin{pmatrix}
		0 & b_1 & & & & \\
		b_1 & 0 & b_2 & & & \\
		& b_2 & 0 & \ddots & & \\
		& & \ddots & \ddots & b_{K-2} & \\
		& & & b_{K-2} & 0 & b_{K-1} \\
		& & & & b_{K-1} & 0
	\end{pmatrix}
\end{align}
where diagonal elements are 0 and off-diagonal elements ($b_1,b_2,\ldots b_n$) form an ordered set of coefficients referred to as Lanczos coefficients. Under time-evolution with $\mathcal{L}$,
\begin{align}
	\mathcal{O}(t)=e^{\imath \mathcal{L}t}|\mathcal{O}_0)=\sum_{n=0}^{K-1} \imath^n \phi_n |\mathcal{O}_n)
\end{align}
where $\phi_n$ are the expansion coefficients of $\mathcal{O}(t)$ in $\mathcal{K}_O$. The unitarity of time evolution ensures $\sum_{n=0}^{K-1}|\phi_n|^2=1$. The time-evolution of $\phi_n$ are given by a set of $K$ coupled differential equations,
\begin{align}
\label{coupled_eqns}
	  \partial_t{\phi}_n(t)= \phi_{n-1}(t)b_{n} - \phi_{n+1}(t)b_{n+1} ,
\end{align}
where $n \in [0,K-1]$ and $\phi_n(0)=\delta_{n,0}$ is taken as initial condition, $b_0=b_{K}=0$ are the boundary conditions. The form of $\mathcal{L}$ suggests an interpretation of this dynamics analogous to a particle hopping problem on a finite chain, with $\{|O_n)\}_{n=0}^{K-1}$ acting as the sites and the hopping amplitudes given by the Lanczos coefficients. To elucidate this, we can represent $\mathcal{L}$ in the Krylov basis,  
\begin{align}
\mathcal{L}=\sum_{n=0}^{K-1}b_{n}(|\mathcal{O}_n)(\mathcal{O}_{n-1}|+|\mathcal{O}_{n-1})(\mathcal{O}_{n}|).
\end{align}
\begin{figure}[htbp]
	\includegraphics[width=\columnwidth]{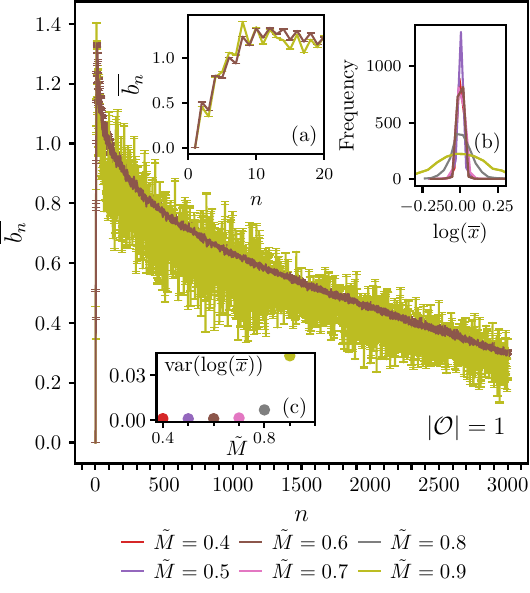}
	\caption{Analysis of the Lanczos coefficients, $b_n$, for an initial operator with support on a single site. The main panel shows the average coefficients, $\overline{b_n}$, as a function of the iteration number $n$ for $\tilde{M}=0.6,0.9$. Error bars represent the standard error of the mean. 
( Inset )(a) Shows the initial growth of $\overline{b_n}$ for the first 20 coefficients. 
(b) Shows the distribution of the logarithm of the ratio of consecutive coefficients, $\ln(\overline{x})$, where $x_n = b_n/b_{n+1}$. 
(c) Shows the variance of the $\ln(\overline{x})$ distribution as a function of $\tilde{M}$. 
All data are for a system of size $L=6$ with Hamiltonian (Eq.~\ref{Hamiltonian}) parameters $g=1.0$ and $h=0.1$, averaged over $500$ graph realizations (denoted by overbar).
  \label{Lanczos_L_6}}
\end{figure}

The corresponding evolution of wavefunction representing this particle, $\mathbb{\phi}=( \phi_0,\imath \phi_1,\ldots,\imath^{K-1}\phi_{K-1})$ can be obtained by solving the set of coupled equations in  \ref{coupled_eqns}. The KC is defined as,
\begin{align}
	\mathcal{C}(t) = \sum_{n=0}^{K-1} n |\phi_n(t)|^2,
\end{align}
 measuring the particle's mean position in the Krylov chain. $|\mathcal{O}_n)$ involves $n$-th order nested commutators, resulting in Krylov basis elements that have increasingly larger support and more complicated structure with increasing $n$. KC captures the spread of a time-evolved operator in this Krylov basis. We denote the support of an operator $\mathcal{O}$ as $|\mathcal{O}|$. 

The Lanczos coefficients $b_n$ control how quickly operators spread through the Krylov basis during time evolution. The scaling of $b_n$ with $n$ and their fluctuations serve as a signature that can distinguish quantum chaotic dynamics from integrable or localized dynamics \cite{trigueros2022krylov}. In the following discussion, we take the initial operator to be in simple product form,
\begin{align}
	\mathcal{O}(0)=\prod_{i=1}^{|\mathcal{O}|}Y_i.
\end{align}
Here, $|\mathcal{O}|$ denotes the size of $\mathcal{O}$, i.e., the number of sites where the operator acts non-trivially. Note that there is no underlying spatial structure in the lattice; thus, no natural local structure can be attributed to $\mathcal{O}(0)$ when it has support larger than $1$. In the rest of the section, we present the empirical observations regarding the Krylov complexity and Lanczos coefficients, along with comparisons with related similar studies and other systems.

\paragraph*{Fluctuation in Lanczos coefficients : } The sequence of Lanczos coefficients $b_n$ has been investigated across different dynamical regimes~\cite{bhattacharya2024krylov,nandy2025quantum}. Empirically, it is found that ${b_n}$ grows linearly (sub-linearly) at small $n$ for chaotic (integrable) systems, saturates at intermediate $n$, and finally descend toward $n=K-1$. The inset (a) in Fig.~\ref{Lanczos_L_6} shows the growth and subsequent saturation of Lanczos coefficients for small $n$. The distinction in the scaling of Lanczos coefficients at small $n$ between chaotic and integrable regimes is unclear here because of the small system size ($L=6$).

The descent region in $b_n$ shows significantly larger fluctuations in the integrable regime than in the chaotic regime. In a single-particle picture, this translates into greater disorder in local hopping parameters. In Fig.~\ref{Lanczos_L_6} (main panel), we show the Lanczos sequence, $\overline{b_n}$ (averaged over ER graphs) for $\tilde{M}=0.6$ and $0.9$.  The fluctuation in Lanczos amplitude at large $n$ is significantly larger for $\tilde{M}=0.9 $ (approaching integrability) compared to $\tilde{M}=0.6 $ (chaotic). Similar large fluctuations were also observed in the sparsely connected regime (not shown).
To quantify this fluctuation, we use the distribution of the ratio of consecutive Lanczos coefficients, $x_n=\left(\frac{b_n}{b_{n+1}}\right)$ averaged over graph realizations. Figure~\ref{Lanczos_L_6} (b) shows the distribution of $\ln(\bar{x}_n)$ with $n$ for multiple $\tilde{M}$. The distribution has a zero mean with the variance increasing as $\tilde{M} \to 1$ (Fig.~\ref{Lanczos_L_6} (c)).

\begin{figure}[htbp]
	\includegraphics[width=\columnwidth]{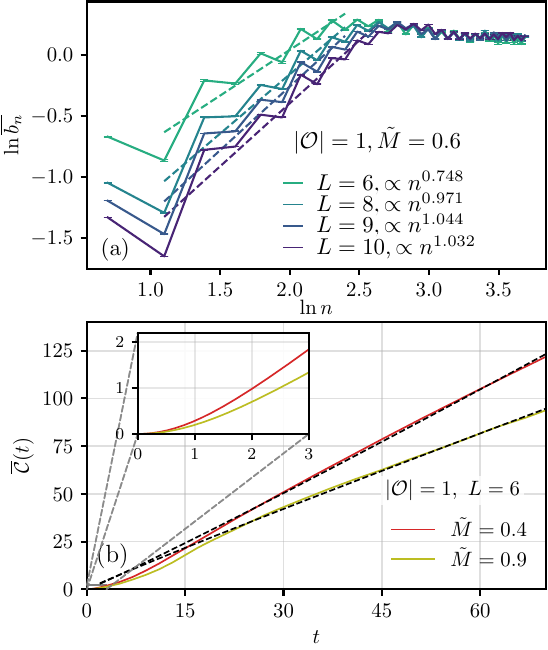}
	\caption{First few Lanczos coefficients and early-time growth of Krylov complexity. 
(a) Log-log plot of the average Lanczos coefficients, $\overline{b_n}$, the iteration number $n$ for several system sizes $L$ at $\tilde{M}=0.6$. Dashed lines show power-law fits, $\overline{b_n} \propto n^\delta$. All data points are averaged over $200$ graph realizations.
(b) The corresponding early-time growth of Krylov complexity, $\overline{\mathcal{C}}(t)$, for the chaotic ($\tilde{M}=0.4$, red) and near-integrable ($\tilde{M}=0.9$, yellow) regimes. The black dashed lines show the linear fit to late-time data. The inset shows a magnified view of the initial non-linear growth.
All data are for a single-site initial operator ($|\mathcal{O}|=1$) with Hamiltonian ( Eq.~\ref{Hamiltonian}) parameters $g=1.0$ and $h=0.1$. For KC, we performed a quenched average over $500$ graph realizations (denoted by overbar).
\label{linear_krylov_complexity_L_6_with_inset}}
\end{figure}

\begin{figure}[htbp]
	\includegraphics[width=\columnwidth]{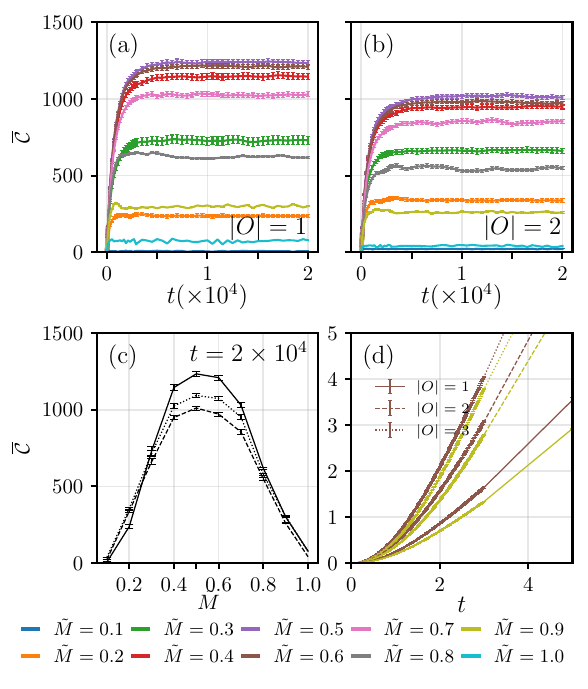}
	\caption{Evolution of Krylov complexity, $\overline{\mathcal{C}}(t)$, for a system with $L=6$. 
(a)-(b) The time evolution for initial operators with increasing support, $|\mathcal{O}|=1$ and $2$ with $\mathcal{O}= \prod\limits^{|\mathcal{O}|}_{i=1}Y_i$, respectively, across multiple connectance, $\tilde{M}$. Error bars indicate the standard error of the mean.
(c) The asymptotic $\overline{\mathcal{C}}$ as a function of $\tilde{M}$ for $|\mathcal{O}|=1, 2$ and $3$ (marked with solid,dashed and dotted black lines, respectively).
(d) A magnified view of the initial growth for the same $|\mathcal{O}|=1, 2$ and $3$ for $\tilde{M}=0.6$ and 0.9. 
All data are presented after averaging over $500$ graph realizations (denoted by overbar) with Hamiltonian  ( Eq.~\ref{Hamiltonian}) parameters $g=1$ and $h=0.1$ from Eq.~\ref{Hamiltonian}.\label{krylov_complexity_L_6}}
\end{figure}

\paragraph*{Growth of KC : }The growth of the Lanczos coefficients, $b_n$, directly governs the multi-stage evolution of the Krylov complexity, $\overline{\mathcal{C}}(t)$. Figure~\ref{linear_krylov_complexity_L_6_with_inset}(a)  shows the $n$ dependence of $\overline{b_n}$ for a single-site initial operator, $\mathcal{O}=Y_1$ in the chaotic regime ($\tilde{M}=0.6$). $\overline{b}_n$ is expected to grow linearly with $n$ in the chaotic regime according to the universal operator growth hypothesis~\cite{parker2019universal,rabinovici2021operator}.
Upon fitting $b_n$ calculated in finite systems to $n^\delta$, we find that $\delta$ approaches $1$ with increasing system size $L$, consistent with chaotic behavior. 
For times $t \lesssim \log(L)$, this linear growth of $b_n$ results in an exponentially growing KC, a feature associated with the spreading of operator support across the system~\cite{rabinovici2022krylov,tang2023operator}. For $L=6$, we find  the initial growth of $\mathcal{C}(t)$ to be slower than this expected exponential behavior (Fig.~\ref{linear_krylov_complexity_L_6_with_inset}(b)), likely related to sublinear growth of $b_n$ with $n$ in smaller systems (Fig.~\ref{linear_krylov_complexity_L_6_with_inset}(a)).

This initial phase is followed by a robust linear growth, indicated by the linear fits (dashed lines) in Fig.~\ref{linear_krylov_complexity_L_6_with_inset}(b), during which the operator spreads through the higher-order elements of the Krylov basis. Finally, at late times, the finite size of the system halts this spread, causing the complexity to saturate. These qualitative features i.e., initial non linear growth that transitions into a linear region in KC, and saturation at late-time because of finite $L$ are present in both the chaotic and near-integrable regimes, though the initial growth rate of KC  slows down near integrability (Fig.~\ref{linear_krylov_complexity_L_6_with_inset}(b) inset).

We observe an even-odd fluctuation about the linear trend in $\overline{b_n}$ similar to what has been observed in quantum field theories and Hamiltonian systems~\cite{avdoshkin2024krylov,rabinovici_localization_2022krylov,trigueros2022krylov,bhattacharjee2022krylov}. The precise form of the subleading corrections to the linear growth relates the long-time decay of the operator autocorrelation. Corrections with an odd-even structure of the form $(-1)^n (\ln n)^{-a}$  have been shown to imply a power-law decay of the autocorrelation~\cite{parker2019universal,bhattacharjee2022krylov}. In the finite data in Fig.~\ref{linear_krylov_complexity_L_6_with_inset}(a), it is unclear if the corrections scale this way.

\paragraph*{Saturation value of KC:} Among the features in the time evolution of KC, the asymptotic (late-time saturation) value $\overline{\mathcal{C}}(t\rightarrow\infty)$, is the clearest indicator of chaos. The saturation value is expected to be higher for chaotic systems compared to integrable or localized systems, reflecting at the emergence of more complex operators at late-times where dynamics are not hindered by constraints or approximate conserved charges. This is evident in Figs.~\ref{krylov_complexity_L_6}(a) and ~\ref{krylov_complexity_L_6}(b), where the KC saturates to a significantly larger value (in the range of $1000-1400$) for intermediate $\tilde{M}$, compared to the extreme $\tilde{M}$ where $\overline{\mathcal{C}}(t\rightarrow\infty) \approx 5-200$ for all operator sizes $|\mathcal{O}_0|$. 
Figure ~\ref{krylov_complexity_L_6}(c) shows the KC value at saturation as a function of $\tilde{M}$, clearly indicating the distinct nature of operator spreading in the chaotic and integrable/localized regimes.
The saturation at a value below the theoretical maximum of $K/2$ can be attributed to finite system size and potential degeneracies in the spectrum. 

\paragraph*{Effect of initial operator size:} Figure~\ref{krylov_complexity_L_6}(d) shows that the initial growth rate increases with the support of the initial operator, $|\mathcal{O}_0|$. This is physically plausible, as an operator with a larger initial support has already ``explored'' a larger part of the system's real space. This feature holds in both chaotic and near-integrable regimes.

We do not observe a simple monotonic dependence of the KC saturation value on the initial operator support $|\mathcal{O}|$, and these results do not change qualitatively for other initial operators, as shown in Appendix~\ref{sec:RandomKrylov}. In contrast, in studies on $\mathrm{SYK}_2$ (a free model), the growth of $\mathcal{C}(t)$ was found to depend on initial operator size~\cite{PhysRevA.105.L010201}.

\begin{figure}[htbp]
	\includegraphics[width=\columnwidth]{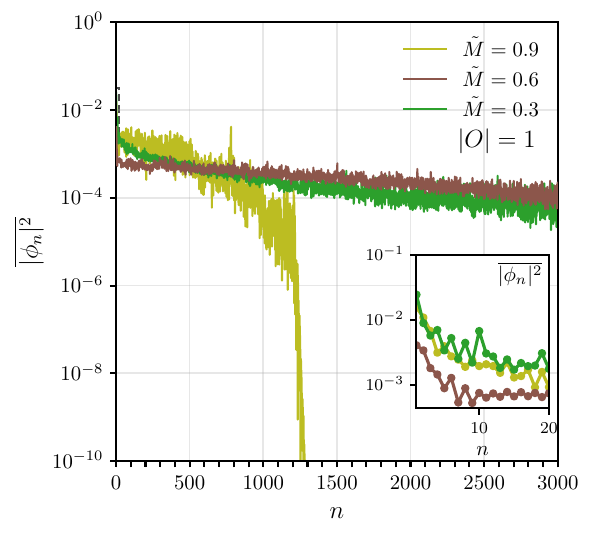}
	\caption{The late-time Krylov wavefunction, $|\phi_n(t)|^2$, shown at $t=2\times10^5$ for several connectance, $\tilde{M}$. (Inset) A magnified view of the occupancy for the first few Krylov basis states at the late-time. All data are for a single-site initial operator ($\mathcal{O}=Y_1$) on a system of size $L=6$ with Hamiltonian parameters $g=1.0$ and $h=0.1$, averaged over $500$ graph realizations (denoted by overbar).  \label{krylov_wf}}
\end{figure}
\paragraph*{Wavefunction spread in Krylov space:} The analogy of a particle hopping on the Krylov chain can be made more direct by examining the probability distribution of the particle's position, given by the Krylov wavefunction, $|\phi_n|^2$. This is reported in Fig.~\ref{krylov_wf} for multiple $\tilde{M}$ at a late-time of $t \approx 2 \times 10^5$. Finite $|\phi_n|^2$ even at large $n$ indicates that $\mathcal{O}(t)$ involves high-complexity operators in the Krylov basis. We observe that $|\phi_n|^2$ decays rapidly with $n$ at large $\tilde{M}$, indicating that the operator remains localized near the beginning of the Krylov chain. In contrast, it shows a much slower decay with $n$ in the chaotic regime, signifying delocalization across the chain. The qualitative trend holds for all operator sizes $|\mathcal{O}|$. 

Experimental access to late-time Krylov complexity is challenging, and the development of scalable experimental protocols is still in its nascent stage~\cite{_indrak_2024,nandy2025quantum}. Thus even though the asymptotic values of Krylov complexity shows clear signatures of chaos as seen in Fig.~\ref{krylov_complexity_L_6} experimentally reproducing them or obtaining a clear scaling behavior is difficult. However early time behavior of Krylov complexity can be captured in current devices by measuring two-point correlator $F(t)=\langle \mathcal{O}(t) \mathcal{O}(0) \rangle$.
Since $F(t)$ is an even function of time (because of the cyclicity of trace), it can be expanded in even powers of $t$ as
\begin{align}
F(t)=\sum_{n=0}^{\infty} \frac{(-1)^n t^{2n}}{(2n)!} \mu_{2n},
\end{align}
where the `moments' $\mu_{2n}$ are related to the adjoint action of the Liouvillian on the operators as $\frac{\rm{Tr}(\mathcal{O}\mathcal{L}^{2n}\mathcal{O})}{\rm{Tr}(\mathcal{O}^2)}$ are the moments of the correlation function. The Lanczos coefficients $\{b_n\}$, which govern the growth of the operator in Krylov space, are entirely determined by these moments via a Gram-Schmidt orthogonalization of the sequence $\{\mathcal{O}, \mathcal{L}\mathcal{O}, \mathcal{L}^2\mathcal{O},\ldots\}$, meaning that $F(t)$ encodes, in principle, all the information needed to extract $\{b_n\}$~\cite{parker2019universal}. Since the polynomial fit requires accurate estimation of the autocorrelation function, which becomes challenging at large $t$ due to decoherence in the experimental setup, only early-time estimates of KC are likely to be accessible. These can still capture certain distinguishing characteristics of chaotic dynamics (Fig.~\ref{linear_krylov_complexity_L_6_with_inset}).

\section{Conclusion and Outlook\label{Discussion_section}}
In this work, we have investigated the emergence of quantum chaos and complexity in the mixed-field Ising model defined on Erd\H{o}s--R\'enyi graphs, and studied how the connectivity $\tilde{M}$ governs the onset of chaos. 
Hamiltonians of similar structure appear naturally in quantum circuits related to optimization tasks, and careful use of such Hamiltonian terms can have a catalytic effect on improving the performance of QAOA and annealing algorithms (Sec.~\ref{QAA_section}, also see Ref.~\cite{Schl_mer_2025} ).
Using a combination of diagnostics -- deep thermalization of the projected ensemble, the partial spectral form factor, and Krylov complexity, we provide a detailed characterization of the localized-to-chaotic-to-integrable crossover driven by changes in connectance. All metrics consistently point to the onset of chaotic behavior at intermediate $\tilde{M}$, even for modest system sizes. With increasing $L$, the chaotic region in the space of $\tilde{M}$ appears to expand for the system sizes studied here, consistent with expectations that in the thermodynamic limit, the model is chaotic everywhere except at $\tilde{M} = 0$ and $1$.

Deep thermalization of the projected ensemble reveals a rapid approach of PE to Haar ensemble with increasing system size at intermediate $\tilde{M}\sim 0.4-0.7$ indicating chaotic dynamics. Away from intermediate $\tilde{M}$, the metrics of deep thermalization, namely the trace distance of PE moments from those of the Haar ensemble, show a smooth crossover from chaotic to integrable/localized features. This is reflected in the time dependence of the trace distances as well as the system size dependence of the asymptotic plateau in the trace distance. The $\tilde{M}\sim 0$ systems have disconnected or weakly connected clusters, which result in local integrals of motion. 
At $\tilde{M}= 1$, i.e., the LMG limit, the system commutes with the total spin operator $S^2$ and also possesses a large number of permutation symmetries, causing fragmentation of the Hilbert space into fragments of sizes at most $L+1$. Away from this limit, the deletion of a finite density of bonds on the LMG model results in weak mixing of these fragments. Thus, just away from the extreme limits, we expect approximate conservation laws that cause slow relaxation. These also cause non-ergodic eigenstates, which are reflected in the asymptotic trace distance plateau being different from that of the maximally chaotic $\tilde{M}\sim 0.4$.
These diagnostics can be scaled up in currently accessible quantum simulators, making our finite-size results a baseline for future experimental efforts. Construction of the projected ensemble, estimation of its higher moments, and behavior of deep thermalization measures have already been demonstrated on quantum simulator platforms with up to $25$ qubits in both neutral-atom arrays~\cite{choi2023preparing} and superconducting architectures~\cite{cotler2023emergent}. The finite-size signatures of deep thermalization reported here, such as the $\tilde{M}$-dependent exponent in the time-dependent decay of the trace distance, are within reach of current devices. With improvements in coherence time, larger system sizes will become accessible on quantum devices, enabling extrapolations toward the thermodynamic limit.

We have shown that the partial spectral form factor also provides a clear characterization of the chaotic dynamics of the random graph Hamiltonian, similar to what has been achieved using the spectral form factor. The intermediate $\tilde{M}$ graphs agree well with the chaos imprints exhibiting a correlation hole, a ramp, and a plateau structure. The localized Hamiltonian in the small $\tilde{M}$ regime has a pSFF without a correlation-hole structure. Both $\tilde{M}\sim 1$ and $\tilde{M}\sim 0$ systems show larger fluctuations and show deviations in the mean value of the late-time plateau of the pSFF. Overall, the behavior of the pSFF in the localizing, integrable, and chaotic regimes is qualitatively similar to that of the spectral form factor. While SFF and pSFF can be estimated using randomized measurements, pSFF requires measurements and controlled unitaries only at a small subsystem level~\cite{joshi2022probing,dong2025measuring}. The better experimental scalability makes the pSFF a feasible tool for investigating the emergence of chaos in larger systems using quantum simulators. Realizing the full ramp-and-plateau structure of the pSFF does, however, require long coherence times to evolve the system to $t \sim t_p$, which scales exponentially with $L$. The correlation hole appears considerably earlier at $t_{\rm Th}$ and, despite its higher measurement overhead, serves as a more practical discriminator between chaotic and nonchaotic regimes in experimentally accessible time windows.

The relaxation dynamics studied in Appendix~\ref{Mpemba_section} also show distinct long-time behavior in the different connectance regimes. While the chaotic regime shows anomalous thermalization behavior reminiscent of the quantum Mpemba effect, the integrable regime shows persistent oscillations in the late-time trace distance from the equilibrium density matrix. 

Krylov complexity provides the cleanest distinction between dynamical regimes, showing clearly larger values at intermediate connectance consistent with chaos (Fig.~\ref{krylov_complexity_L_6}(c)). However, estimating KC experimentally is challenging at present. Experimentally measurable proxies for KC have been proposed; these can help scale up the study of the emergence of chaos from an information scrambling perspective.
An analogous investigation applicable directly to the quantum annealing setting requires analysis of the low-energy sector of the spectrum using probes such as an energy-density-specific Krylov complexity, as the performance of quantum annealing is sensitive to the localization of low-lying eigenstates~\cite{cain2023quantum}. 
Overall, our results provide a comprehensive and experimentally relevant characterization of the connectivity-driven onset of quantum chaos, offering benchmarks and insights for controlling the performance of variational quantum algorithms and reservoir computing~\cite {moein_reservoir}.

\begin{acknowledgments}
We thank S Roy for useful discussions and collaborations
on an earlier related work. We thank National Supercomputing Mission (NSM) for providing computing resources of ``PARAM Brahma" at IISER Pune, which is implemented by C-DAC and supported by the Ministry of Electronics and Information Technology (MeitY) and Department of Science and Technology (DST), Government of India.
\end{acknowledgments}

\clearpage
\appendix
\section{Eigenstate Localization and Entanglement \label{ipr_appendix}}
\begin{figure}[h]
	\includegraphics[width=\columnwidth]{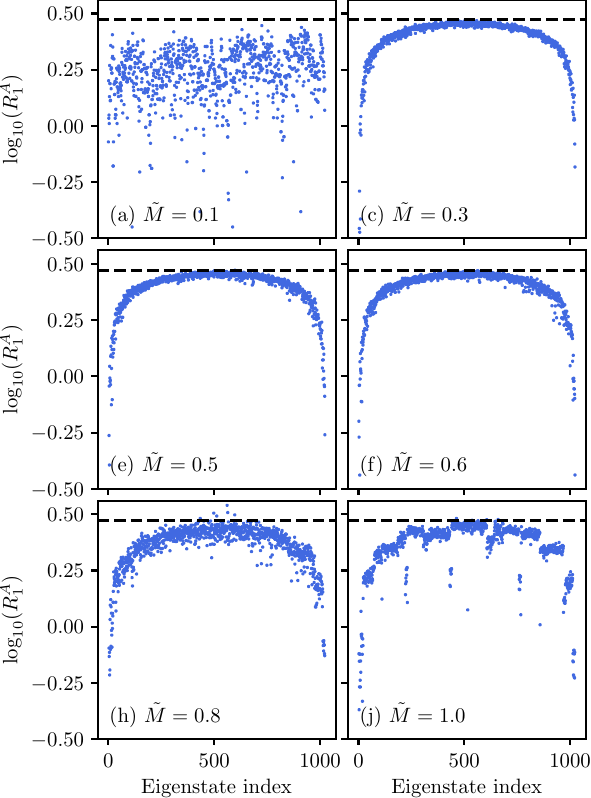}
	\caption{ The bipartite entanglement entropy for a single instance of the Ising problem on an ER graph of $L=10$ sites with $L_A=5$. The black dashed line shows the Page value.  \label{entanglement_eigesnstate}}
\end{figure}

Here, we briefly discuss the eigenstate properties of the Hamiltonian in Eq.~\ref{Hamiltonian} with $g=1$ and $h=0.1$ for a graph with $L=10$. We take a single realization of the Hamiltonian at each $\tilde{M}$ and compute the bipartite entanglement entropy of each eigenstate. We quantify the entanglement entropy with Von-Neumann entropy, $R_1^A=-\sum_i \lambda_i \ln \lambda_i$, where $\lambda_i$ are the squared singular values of the subsystem reduced density matrix of the subsystem. The results are presented in Fig.~\ref{entanglement_eigesnstate} for a range of connectivities. The maximum entanglement is bounded by the Page value, $R_1^{\rm {A,Page}}=(L/2) \ln 2-0.5$ and is marked with the black dashed line. The entanglement is smaller than Page value even for the mid-spectrum states at $\tilde{M}$ close to $0$. For $\tilde{M}\approx 0.3-0.7$, the entanglement entropy of most of the bulk eigenstates approaches the Page value. As $\tilde{M}\to 1.0$, a fraction of mid-spectrum eigenstates exhibits lower entanglement entropy. At $\tilde{M}= 1.0$, signs of fragmentation appear, with the eigenstate entanglement values clustering.

\begin{figure}[h!]
	\includegraphics[width=\columnwidth]{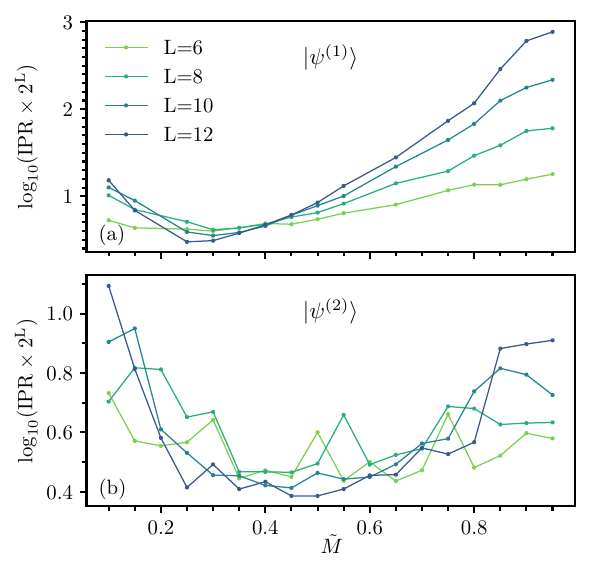}
	\caption{ The IPR of product states (a) $|\psi^{(1)} \rangle$ (b) $|\psi^{(2)} \rangle$. All data points are averaged over $100$ graph instances (and initial state in the case of (b))  \label{ipr_eigesnstate}}
\end{figure}
Next, we examine the localization of simple product states in the eigenstate basis. We measure this quantity with inverse participation ratio (IPR), defined as
\begin{align}
\mathrm{IPR}(\psi)=\sum_{i=1}^d |\langle E_i | \psi\rangle|^4,
\end{align}
where $E_i$ are Hamiltonian eigenstates. IPR is $1$ for perfectly localized state and $1/2^L$ for maximally delocalized states. We compute the IPR for two classes of initial product states,
\begin{align}
    |\psi^{(1)} \rangle &= \frac{1}{\sqrt{2^L}}\bigotimes_{i=0}^L ( |0 \rangle + \imath |1\rangle), 
\end{align}
These states (the same as those in Eq.~\ref{projected_initstate}) have zero energy and are permutation-invariant.  Next, we consider a class of states that lack the permutation invariance. Consider that each site $i$ is prepared in an equatorial state on the Bloch sphere,
 \begin{equation}
    |\psi^{(2)}\rangle = \bigotimes_{i=1}^{L} |\phi_i\rangle, 
    \qquad 
    |\phi_i\rangle = \frac{|0\rangle + e^{i\phi_i}|1\rangle}{\sqrt{2}},
    \label{eq:initial_state_PE_random}
\end{equation}
where $|0\rangle$ and $|1\rangle$ are the eigenstates of $Z$. The angles
$\{\phi_i\}$ are drawn from antipodal pairs to ensure zero energy. Specifically, $L/2$ angles $\alpha_k \sim \mathrm{Uniform}[0, 2\pi)$ are sampled independently, giving the set
 \begin{equation}
    \{\phi_i\}_{i=1}^{L} 
    = \{
        \alpha_1,\, \alpha_1 + \pi,\;
        \alpha_2,\, \alpha_2 + \pi,\;
        \ldots,\;
        \alpha_{L/2},\, \alpha_{L/2} + \pi
    \},
    \label{eq:angle_set}
\end{equation}
randomly assigned across sites. The zero-energy condition follows from the cancellation within each antipodal pair,
\begin{equation}
    \langle H \rangle 
    = -g \sum_{i=1}^{L} \cos\phi_i 
    = -g \sum_{k=1}^{L/2} \!\Big(\cos\alpha_k + \cos(\alpha_k + \pi)\Big) 
    = 0,
    \label{eq:zero_energy}
\end{equation}
where the $J_z$ and $h_z$ terms vanish identically since $\langle \sigma^z_i \rangle = 0$ for all equatorial states. The special case $\alpha_k = \pi/2 $ for all $k$ recovers the uniform $|+ y\rangle^{\otimes L}$ initial state used in the main text in Eq.~\ref{projected_initstate}.
 Figure~\ref{ipr_eigesnstate}(a) shows the IPR for $|\psi^{(1)} \rangle$ averaged over $100$ Hamiltonians. The IPR is minimum near $\tilde{M}=0.3$, indicating the state is maximally spread among the eigenstates compared to the case of other $\tilde{M}$. Incidentally, $\Delta^{(k)}$ attains its minimum value around this $\tilde{M}$ in Fig.~\ref{asymptotic_connectance}. The IPR goes up (the product state becomes localized in the eigenbasis of the Hamiltonian) at extremal values of $\tilde{M}$. 

For $|\psi^{(2)} \rangle$, we average over both initial state and Hamiltonians at each $\tilde{M}$. As shown in Fig.~\ref{ipr_eigesnstate}(b), for larger systems, there is a clear minimum in IPR at $\tilde{M}=0.5$. At very small systems ($L=6$), this distinction becomes weak.

\section{Additional data on QAOA with random Ising mixer \label{qaoa_appendix}}
\begin{figure}[]
	\includegraphics[width=\columnwidth]{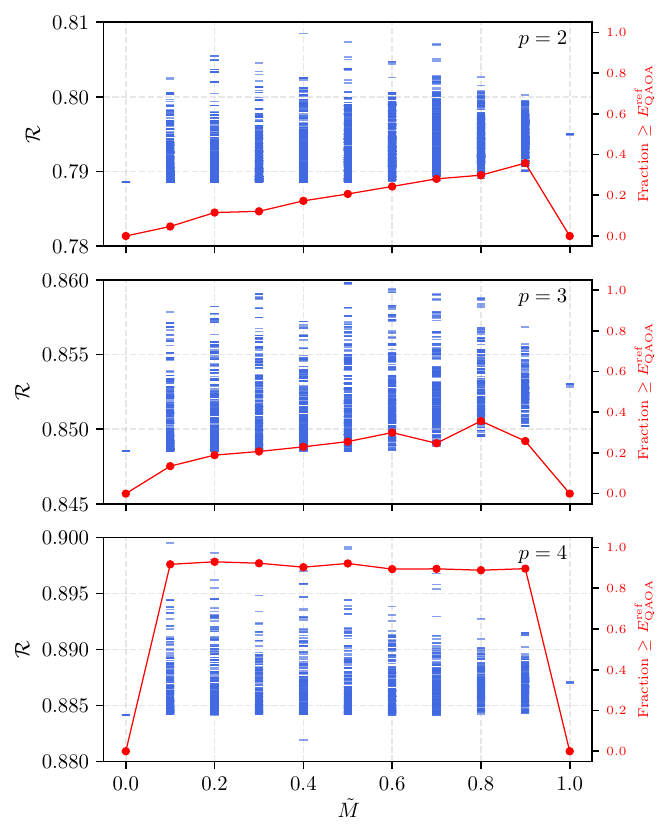}
	\caption{ The approximation ratio for an Ising problem on an ER graph of $L=10$ sites with $\tilde{M}_c=0.3$. The red line shows the fraction of QAOA instances where the corresponding connectivity performs better than the baseline.  \label{qaoa_10}}
\end{figure}

In Sec.~\ref{QAOA}, we provided an illustrative example demonstrating the effect of the additional mixer Hamiltonian on QAOA performance  for $L=8$ and $\tilde{M}_c=0.4$. Here, we present the QAOA performance for an random Ising problem instance on an ER graph with $L=10$ and $\tilde{M}_c=0.3$ to validate the results on a larger system. The $H_d$ was taken with Pauli-$YY$ coupling (Eq.~\ref{qaoa_Hd}). This $YY$ term can scatter between computational basis states with Hamming distance $2$ compared to just the transverse field mixer which scatters among states with Hamming distance $1$. We present the numerical results in Fig.~\ref{qaoa_10} for $500$ random instances of $H_d$ at each $\tilde{M}$. Here, we first define a baseline for performance comparison. We consider the best $E_{\text{QAOA}}$ obtained with $\tilde{M} \in \{0,1\}$ as $E^{\text{ref}}_{\text{QAOA}}$. The red line denotes the fraction of instances for which the energy obtained with QAOA protocol is lower compared to $E^{\text{ref}}_{\text{QAOA}}$. The fraction of instances surpassing the baseline energy shows two distinct trends: at fixed $p$, it is maximized at intermediate $\tilde{M}$, confirming that the chaotic regime provides a favorable landscape for QAOA optimization; additionally, at any fixed $~\tilde{M}$, this fraction grows with $p$, indicating that deeper circuits with more applications of the chaotic driver Hamiltonian progressively improve the ability to explore the energy landscape and find lower-energy solutions. The latter trend is consistent with the expectation that repeated application of a chaotic driver generates more complex quantum superpositions, allowing QAOA to access a larger portion of the Hilbert space with increasing circuit depth. This trend empirically holds for circuits with up to $p=4$ layers, which are the ddepest circuits studied in this work. However, very deep quantum circuits suffer from barren plateaus where the cost function gradient vanishes exponentially with system size, making classical parameter optimization intractable.

\begin{figure}[h]
	\includegraphics[width=\columnwidth]{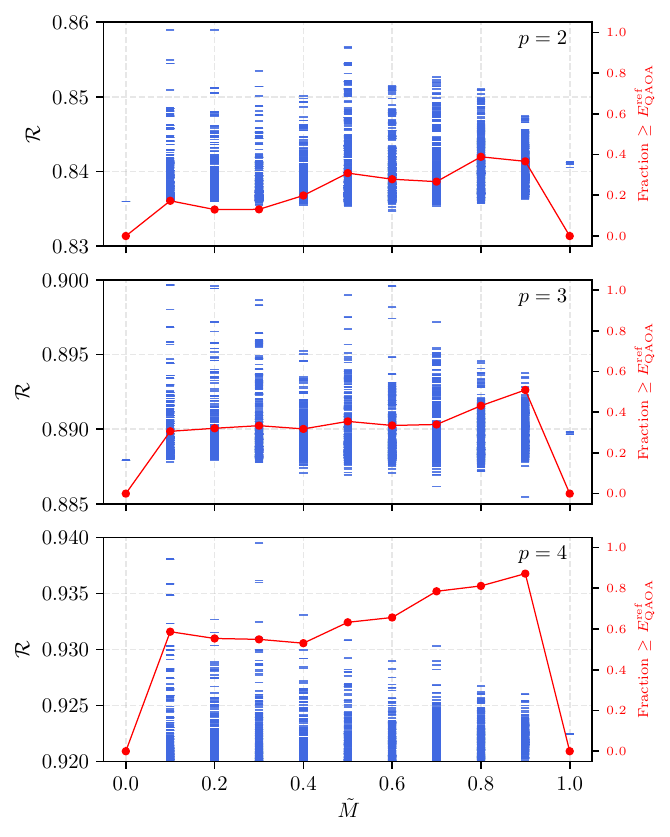}
	\caption{ The approximation ratio for an Ising problem on an ER graph of $L=8$ sites with $\tilde{M}_c=0.4$ (same $H_{\rm c}$ as Problem 1 in Fig.~\ref{qaoa_scatter}). The red line shows the fraction of QAOA instances where the corresponding connectivity performs better than the baseline.  \label{qaoa_ZZ}}
\end{figure}

As discussed in Sec .~\ref {QAA_section}, the improvement in QAOA performance observed with the $H^{YY}$ driver could stem from the chaotic nature of $H_{\rm d}$, or from the presence of terms that scatter among computational basis states more efficiently.  We consider an alternative driver Hamiltonian with Pauli-$ZZ$ coupling,

\begin{align}
\label{qaoa_Hd_z}
    H_{\rm d} &= -\frac{JL}{M} \sum_{i,j}A_{ij} Z_iZ_j -g\sum_i X_i,
\end{align}
which is diagonal in the computational basis and therefore induces no scattering among basis states. Taking the same problem instance $H_{\rm c}$ as in Fig.~\ref{qaoa_scatter}(a),(c) (Problem 1), we present the distribution of QAOA performance for 
$p=2,3$ and $4$ in Fig.~\ref{qaoa_ZZ}. The improvement in performance with intermediate $\tilde{M}$
persists with the $ZZ$ driver, despite the absence of any Hamming distance $2$ scattering. This establishes that the chaotic structure of $H_{\rm d}$, rather than the specific scattering properties of the mixer, is the primary driver of the observed improvement in QAOA.

\section{PE from a Permutation-Symmetry-Breaking Initial State\label{sec:PE_random}}

\begin{figure*}[htbp]
    \includegraphics[width=\textwidth]{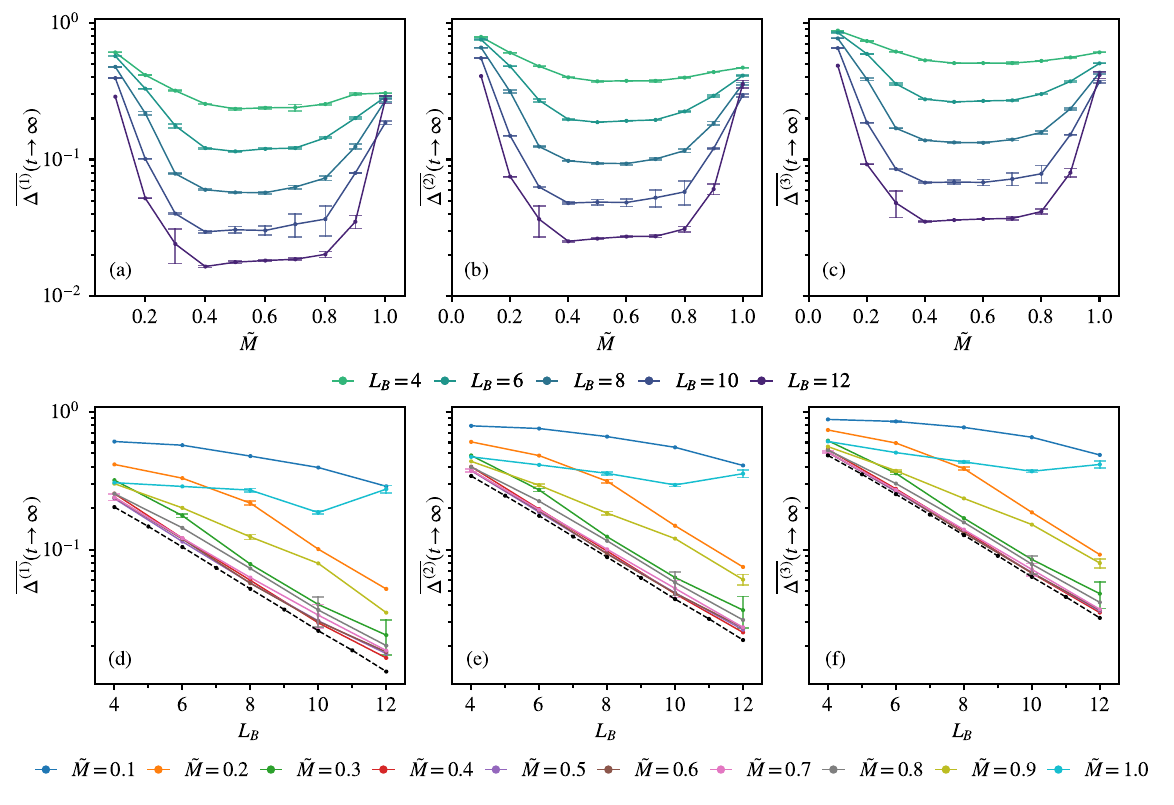}
    \caption{(a)-(c) Asymptotic ($t=10^9J^{-1}$) trace distance of first three moments of the PE from the Haar ensemble as a function of $L_B$. The asymptotic trace distance decreases with increasing $L_B$ for all $\tilde{M}$. Error bars represent the standard error of the mean across realizations at a fixed $t$. (d)-(f) The same data plotted to show the scaling of $\overline{\Delta^{(k)}}(t \to \infty)$ with $L$ across all $\tilde{M}$ for first three moments of the PE. The error bars are not shown when they are comparable to the marker size. Each data point shown represents an average over 500 graph realizations and random direct product initial states (Eq.~\ref{eq:initial_state_PE_random}), calculated for a subsystem of size $L_A=2$ with Hamiltonian parameters (Eq.~\ref{Hamiltonian}) $g=1.0$ and $h=0.1$. The black dashed line shows the trace distance of an empirical ensemble of $500$ random states (sampled from the Haar random states for a system of size $L$) from the corresponding moments of the Haar ensemble. \label{random_asymptotic_connectance}}
\end{figure*}

The initial state is chosen to be the product state $\psi^{(2)}$ (Eq.~\ref{eq:initial_state_PE_random} with zero energy expectation value, $\langle \psi_0 | H | \psi_0 \rangle = 0$ and without permutation symmetry. 

Figure~\ref{random_asymptotic_connectance} shows the scaling of the trace distance between the first three moments of the PE and the Haar ensemble across connectivity and system size. The initial state is taken as a random direct product state (Eq.~\ref{eq:initial_state_PE_random}) with zero energy expectation value. The data are presented for $500$ random initial states and ER graphs. The qualitative behavior is similar to that shown in Fig.~\ref{asymptotic_connectance}.

\section{Lanczos algorithm\label{sec:Lanczos}}
In the main text, we used the Lanczos algorithm to generate a basis for the Krylov space. The Lanczos algorithm is an iterative method for constructing an orthonormal basis of the Krylov subspace, $\mathcal{K}_\mathcal{O}$. The conventional implementation of the algorithm relies on a computationally efficient three-term recurrence relation, in which each new vector is made orthogonal only to the two preceding vectors. While theoretically exact, this procedure is numerically unstable when implemented with finite-precision arithmetic, as rounding errors accumulate, gradually leading to a loss of orthogonality among the basis vectors.

To circumvent this issue, we employ the more robust, albeit computationally intensive, Lanczos algorithm with full orthogonalization~\cite{golub2013matrix,rabinovici2021operator}. In this scheme, each new candidate vector is explicitly made orthogonal to \textit{all} previously generated basis vectors using a Gram-Schmidt procedure. This ensures a numerically stable orthonormal basis (up to machine precision) at the cost of increased computational complexity, as the number of required inner products scales with the iteration number $n$, rather than being constant. However, for small system sizes, this cost is manageable. The procedure we implement is as follows:

\begin{enumerate}
    \item \textbf{Initialization:} Start with the vectorized initial operator, $|\tilde{\mathcal{O}_0}) = \mathrm{vec}(\tilde{\mathcal{O}_0})$. Normalize it to obtain the first basis vector of the Krylov basis:
    \[
        |\mathcal{O}_0) = \frac{|\tilde{\mathcal{O}_0})}{\sqrt{(\tilde{\mathcal{O}_0}|\tilde{\mathcal{O}_0})}}
    \]
    
    \item \textbf{Iteration (for n = 0, 1, 2, ...):}
    
    First, generate a new candidate vector by applying the Liouvillian to the most recently generated basis vector:
    \[
        |w_{n+1}) = \mathcal{L} |\mathcal{O}_n)
    \]
    Next, enforce orthogonality. To ensure the new vector is orthogonal to all previously found basis vectors, explicitly subtract the projection onto each of them. For numerical stability, this process is repeated twice (double orthogonalization):
    \[
        |w'_{n+1}) = |w_{n+1}) - \sum_{j=0}^{n} (\mathcal{O}_j|w_{n+1}) |\mathcal{O}_j)
    \]
    The next Lanczos coefficient, $b_{n+1}$, is the norm of this fully orthogonalized vector:
    \[
        b_{n+1} = \sqrt{(w'_{n+1}|w'_{n+1})}
    \]
    If $b_{n+1}$ is smaller than a numerical tolerance (e.g., $10^{-13}$), the Krylov subspace is exhausted, and the algorithm terminates constructing a Krylov subspace of dimension $K$ such that $b_K < 10^{-13}$. Otherwise, the next orthonormal basis vector is found by normalizing:
    \[
        |\mathcal{O}_{n+1}) = \frac{|w'_{n+1})}{b_{n+1}}
    \]
    
\end{enumerate}
This iterative process yields the set of orthonormal basis operators $\{|O_n))\}_{n=0}^{K-1}$ and the real, positive Lanczos coefficients $\{b_n\}_{n=1}^{K-1}$ that define the tri-diagonal hermitian matrix representation of $\mathcal{L}$.

A key feature of the resulting tridiagonal matrix representation of $\mathcal{L}$ is that its diagonal elements, $a_n = (\mathcal{O}_n| \mathcal{L} |\mathcal{O}_n)$, are identically zero. This arises from the definition of the Liouvillian and the cyclic property of the trace. Assuming the basis operators $|\mathcal{O}_n))$ are Hermitian (which is the case if the initial operator is Hermitian), we have:
\begin{align*}
    a_n &= (\mathcal{O}_n| \mathcal{L} |\mathcal{O}_n) \\
        &= \mathrm{Tr}(\mathcal{O}_n^\dagger [H, \mathcal{O}_n]) \\
        &= \mathrm{Tr}(\mathcal{O}_n (H \mathcal{O}_n - \mathcal{O}_n H) \\
        &= \mathrm{Tr}(\mathcal{O}_n H \mathcal{O}_n) - \mathrm{Tr}(\mathcal{O}_n \mathcal{O}_n H) =0 
\end{align*}

This cancellation ensures that the Liouvillian is purely off-diagonal in the Krylov basis, simplifying the resulting dynamics to a hopping problem on a chain without any on-site potential.

\section{Krylov complexity for random single-site operators\label{sec:RandomKrylov}}

\begin{figure}[htbp]
	\includegraphics[width=\columnwidth]{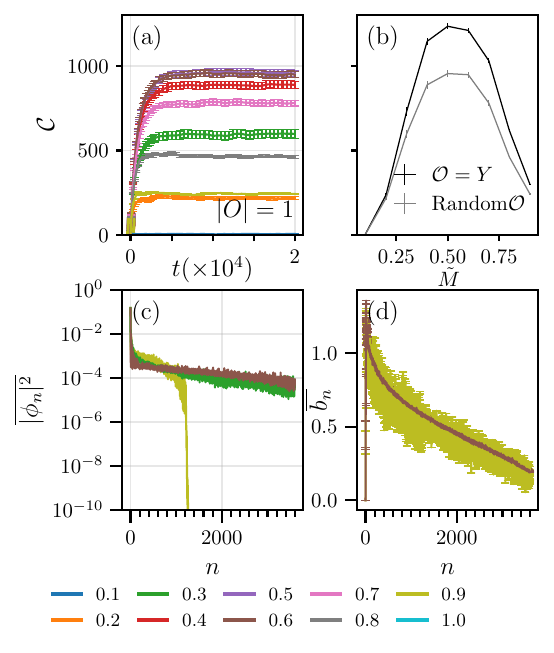}
	\caption{(a) Time evolution of KC and its asymptotic value across multiple $\tilde{M}$. (b) Comparison of asymptotic Krylov complexity as a function of $\tilde{M}$ for random single-site operator and Pauli-$Y$ operator. (c) The asymptotic Krylov wavevector (d) The distribution of Lanczos coefficients. $L=6$ Error bars are shown as the standard error of the mean. All results are presented after averaging over $500$ random graphs and operators. \label{random_krylov_complexity}}
\end{figure}

In Sec.~\ref{Krylov_section}, we discussed the Krylov complexity for a simple initial operator (Pauli $Y$). We also examined the effect of the initial operator size on the growth of the Krylov complexity. A natural question arises regarding the dependence of the behavior on the choice of operator. Here, we take a random single-site operator as,
\begin{align}
    \mathcal{O}_i=r_0 \mathbb{I}+r_1 X_i+r_2 Y_i+r_3 Z_i.
\end{align}
Here, $\{r_0,r_1,r_2,r_3\}$ are random numbers sampled from a normal distribution. The operator is normalized such that it has a norm of $\sqrt{2}$ (i.e., equal to that of the Pauli operators). In Fig~\ref{random_krylov_complexity}, we present the results after averaging over $500$ graphs. For each graph realization, we take a different random initial operator. Averaging is done over both graph realizations and the initial operator. The trend with $\tilde{M}$ holds here as well as shown in Fig~\ref{random_krylov_complexity}(a). We compare the asymptotic KC obtained after averaging over random graphs and operators with that obtained when averaging is performed only over graphs in Fig.~\ref{random_krylov_complexity}(b). The asymptotic KC is smaller for random operators, but the qualitative trend is the same. The time-evolved Krylov wavefunction is shown in Fig.~\ref{random_krylov_complexity}(c). The wavefunction shows a clear sign of localization for $\tilde{M}=0.9$. The distribution of Lanczos coefficients is also consistent with the qualitative characteristics observed earlier in Sec.~\ref{Krylov_section}.

\section{Anomalous relaxation of initial product states \label{Mpemba_section}}
In this Appendix, we analyze the dynamical relaxation of quantum states under unitary time evolution. In closed quantum many-body systems, a subsystem's approach to equilibrium is driven by the scrambling of local information, a process whose efficiency was characterized in Sec.~\ref {Krylov_section} via Krylov complexity. Here, we focus on how certain simple initial states relax under this unitary dynamics. An intuitive expectation for individual states is that the time to reach equilibrium should monotonically increase with the initial state's \quotes{distance} from the equilibrium ensemble. In a classical context, a \quotes{hotter} system is farther from ambient temperature compared to a \quotes{cooler} system. So, we can expect that a \quotes{cooler} state, being \quotes{closer} to equilibrium, would relax faster than a \quotes{hotter} one. However, it has been found that certain initial conditions can defy this intuition, giving rise to anomalous thermalization behaviors. Broadly, this phenomenon is known as the Mpemba effect, first reported in water~\cite{EB_Mpemba_1969}, where initially hotter systems freeze faster than cooler ones under a quench. 

Recently, the quantum Mpemba effect, a quantum analog of the counterintuitive classical phenomenon, has been studied. The QME was first proposed for open quantum systems, where a system's coupling to a thermal bath induces irreversible dynamics~\cite{Lu_2017,Nava_2019,Carollo_2021,Chatterjee_2023}. In this setting, the effect is understood to arise from the spectral properties of the system's dissipative  Liouvillian. An initial state can be designed such that its overlap with the slowest eigenmode of the Liouvillian is suppressed, allowing it to equilibrate anomalously fast. This open-system QME was recently demonstrated experimentally~\cite{Zhang_2025}. 

Subsequently, the QME has been extended to isolated systems with global symmetry undergoing unitary evolution, where thermalization is observed at the subsystem level~\cite{Ares2024SimplerProbe, Rylands2024MicroscopicOrigin, Ares2025QuantumMpemba, Zhu2025SymmetryQME, Bhore2025QuantumMpemba}. In such closed systems, the \quotes{distance} from equilibrium is typically inferred from the off-diagonal blocks in the symmetry-imposed subsystem reduced density matrix at equilibrium after a quench from an initial state which is not an eigenstate of the global symmetry. Recently, QME has been studied in spin chains and Floquet systems where global symmetries are absent~\cite{Bhore2025QuantumMpemba} as well as systems with disorder~\cite{liu2024quantum}.
\begin{figure}[htbp]
	\includegraphics[width=\columnwidth]{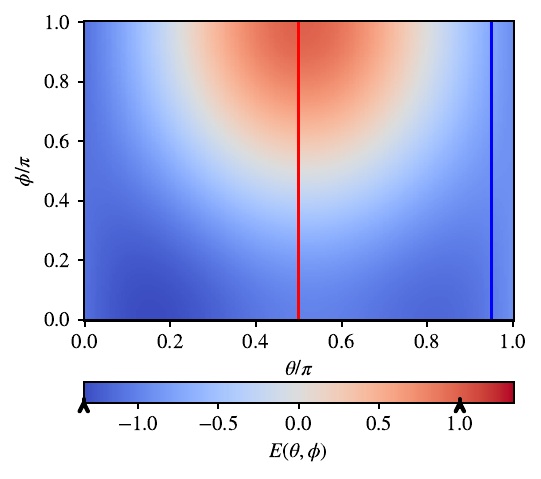}
	\caption{Distribution of energy of the initial state as a function of $(\theta,\phi)$. $L_A=4,L=12$. The initial states are of the form in Eq.~\ref{initstate_mpemba}. The markers on the colorbar indicate the bandwidth of the eigen spectrum. \label{init_mpemba_energy}}
\end{figure}

To investigate the presence of the QME in this random graph system, we first establish a quench protocol and a metric to quantify the thermalization. Unlike studies that track the restoration of a global symmetry, our Hamiltonian (Eq.~\ref{Hamiltonian}) with $ h\neq 0$ lacks such a conserved quantity. We therefore adopt a more direct approach~\cite{Bhore2025QuantumMpemba} to quantify relaxation. Our metric is the trace distance, $\Delta^{(1)}$ (Eq.~\ref{tr_dist_eq}), between the time-evolved reduced density matrix and its corresponding diagonal ensemble. This diagonal ensemble represents the infinite-time equilibrium ensemble ($\rho_A(\infty)$) given as,  
\begin{equation}
	\rho_A(\infty)=\mathrm{Tr}_{B}(\mathcal{U}\mathrm{diag}(|\mathcal{U} \psi \rangle|^2 )\mathcal{U}^{\dagger}),
\end{equation}
where $H=\mathcal{U}\tilde{E}U^{\dagger}$ for Hamiltonian $H$ and $\tilde{E}$ is a diagonal matrix containing the eigenvalues of $H$. $B$ denotes the complement of subsystem $A$. To simplify the parameter space of the initial state, we prepare a family of permutation-invariant product states $\psi(\theta,\phi)$ in real space as,
\begin{align}
    \label{initstate_mpemba}
	\psi(\theta,\phi)= \otimes_{i=0}^L \left(\cos\left(\frac{\theta}{2}\right) |0\rangle +e^{\imath \phi }\sin\left(\frac{\theta}{2}\right) |1\rangle \right).
\end{align}
The initial state energy $E(\theta,\phi)$ is a smooth function of $\theta,  \phi$ and given as,
\begin{equation}
    \label{energy_mpmeba}
	E(\theta,\phi)=-(\cos\theta)^2-g \sin \theta \cos\phi -h\cos \theta.
\end{equation}
$E(\theta,\phi)$ depends only on the coupling parameters of the Hamiltonian but not on the connectance (This independence arise from the normalization factors in the denominator of Eq.~\ref{Hamiltonian}). The distribution of initial state energy as a function of $(\theta,\phi)$ is shown in Fig.~\ref{init_mpemba_energy}. The red and blue regions representhigh and low energy states, respectively, with a crossover region near $0$ energy where states correspond to infinite temperature. Consider two initial states $\psi(\theta_1,\phi_1)$ and $\psi(\theta_2,\phi_2)$ such that $\Delta^{(1)}_{\theta_1,\phi_1}(t=0) > \Delta^{(1)}_{\theta_2,\phi_2}(t=0)$ where trace distance is always taken from the respective diagonal ensembles. The existence of the QME implies that for $t > t_M$, $\Delta^{(1)}_{\theta_1,\phi_1}(t) < \Delta^{(1)}_{\theta_2,\phi_2}(t)$, where $t_M$ is referred to as the Mpemba time. Here, $\Delta^{(1)}_{\theta_1,\phi_1}(t=0)$ is the initial \quotes{hotter} state caused by the higher trace distance of the subsystem reduced density matrix from the corresponding equilibrium ensemble. 

Our analysis focuses on two distinct classes of initial states (Eq.~\ref{initstate_mpemba}), parametrized by $\theta$ and marked in Fig.~\ref{init_mpemba_energy}. The first class, with $\theta=0.5\pi$, spans a wide range of initial energies as $\phi$ is varied. In contrast, the second class, with $\theta=0.95\pi$, consists of nearly-polarized states whose initial energies are only weakly dependent of $\phi$.

\begin{figure}[htbp]
	\includegraphics[width=\columnwidth]{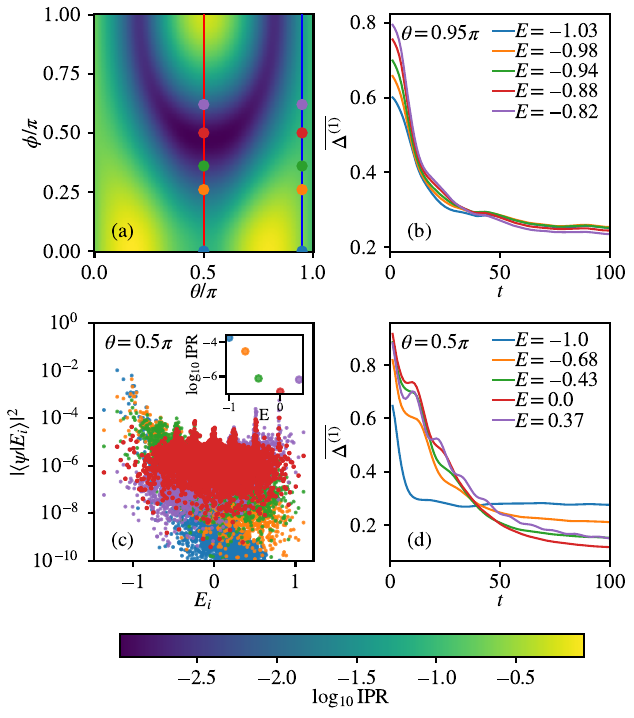}
    \caption{Anomalous relaxation in the chaotic regime at $\tilde{M}=0.4$ for a system with $L=12$. (a) The Inverse Participation Ratio (IPR) as a function of the initial state parameters $(\theta, \phi)$. The red and blue vertical lines at $\theta=0.5\pi$ and $\theta=0.95\pi$ indicate the two classes of states studied. (b) Relaxation dynamics, measured by the trace distance $\Delta^{(1)}(t)$ from the diagonal ensemble, for the nearly-polarized states ($\theta=0.95\pi$). (c) The distribution of overlaps, $|\langle E_n|\psi\rangle|^2$, with energy eigenstates for several initial states from the $\theta=0.5\pi$ class. The inset shows the corresponding IPR for each state. (d) Relaxation dynamics for the $\theta=0.5\pi$ class, showing a clear crossing indicative of the QME. For all plots, the subsystem size is $L_A=4$, and all data are averaged over $500$ graph realizations (denoted by overbar) using the Hamiltonian in Eq.~\ref{Hamiltonian} with $g=1.0$ and $h=0.1$. \label{mpemba_timeevo_0p4}}
\end{figure}

\begin{figure}[h!]
	\includegraphics[width=\columnwidth]{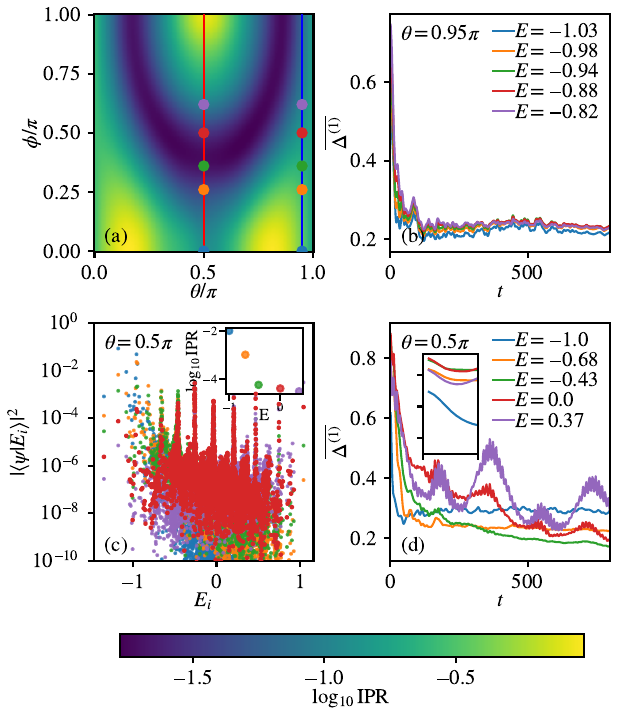}
    \caption{Anomalous relaxation in the near-integrable regime at $\tilde{M}=0.8$ for a system with $L=12$. (a) The IPR as a function of the initial state parameters $(\theta, \phi)$. The red and blue vertical lines indicate the two classes of states studied. (b) Relaxation dynamics, $\Delta^{(1)}(t)$, for the nearly-polarized states ($\theta=0.95\pi$). (c) The distribution of overlaps, $|\langle E_n|\psi\rangle|^2$, with energy eigenstates for several initial states from the $\theta=0.5\pi$ class. (d) Relaxation dynamics for the $\theta=0.5\pi$ class, showing persistent oscillations and multiple crossings. The inset shows a zoomed-in view of the initial relaxation $(t=0-10 (J))$. For all plots, the subsystem size is $L_A=4$, and all data are averaged over 500 graph realizations (denoted by an overbar) using the Hamiltonian in Eq.~\ref{Hamiltonian} with $g=1.0$ and $h=0.1$.\label{mpemba_timeevo_0p8}}
\end{figure}

Starting with these two classes of initial states (distinguished by $\theta$), we time evolve under a specific graph realization of Hamiltonian in Eq.~\ref{Hamiltonian} with $h=0.1,g=1$. We first consider $\tilde{M}=0.4$, where the model is expected to be chaotic. Figure~\ref{mpemba_timeevo_0p4}(b) and (d) shows time evolution of $\overline{\Delta^{(1)}}$ for $\theta=0.5\pi$ and $\theta=0.95\pi$, respectively. All data are presented after averaging over $500$ graph realizations starting with the same set of initial states. We observe a clear crossing in $\Delta^{(1)}$ in Fig.~\ref{mpemba_timeevo_0p4}(d) where the initial states are characterized by $\theta=0.5\pi$. Such a crossing is absent in Fig.~\ref{mpemba_timeevo_0p4}(b) where $\theta=0.95\pi$ and relaxation to equilibrium appears almost independent of the initial state at the late-time. 

The observed differences in the late-time relaxation dynamics can be qualitatively understood by considering how the initial states are represented in the energy eigenbasis. $\Delta^{(1)}$ can be expressed in terms of the overlap of the initial state with energy eigenstates as,
\begin{align}
\label{tr_dist_mpemba}
\Delta^{(1)}(t) = \frac{1}{2} \left\| \sum_{n \neq m} c_n c_m^* e^{-i(E_n - E_m)t} \mathrm{Tr}_{\mathrm{B}}(|E_n\rangle\langle E_m|) \right\|_1
\end{align}
where $c_n=\langle \psi | E_n \rangle$. At $t=0$, $\Delta^{(1)}(0) = \frac{1}{2} \left\| \sum_{n \neq m} c_n c_m^* \mathrm{Tr}_{\mathrm{B}}(|E_n\rangle\langle E_m|) \right\|_1$ i.e., the trace distance contains $O(d^2)$ terms weighted by the overlaps of initial state with eigenstates. As pointed out in Ref.~\cite{Bhore2025QuantumMpemba}, the relaxation rate of states can be related to the number of nonzero coefficients $c_n$, quantified by the inverse participation ratio (IPR) given as,
\begin{align}
\mathrm{IPR}(\psi)=\sum_{i=1}^d |\langle E_i | \psi\rangle|^4.
\end{align}
If the initial state $\psi$ is a generic state  (with energy close to $0$, i.e., analogous to a mid-spectrum eigenstate), $c_n \approx  O(1/\sqrt{d})$ in Eq.~\ref{tr_dist_mpemba},  corresponding to a smaller IPR. 

 $|\Delta^{(1)}(t)|^2$ is the sum of squared norms of the matrix elements of $(\rho_A-\rho_A{(\infty}))$. For an initial state $\psi$ that has overlap with a large number of energy eigenstates (i.e., smaller IPR), there is a large number of nonzero $c_n$s. Each $c_n$ is small because of the normalization of the initial state and, at the same time, there are a larger number of fluctuating phases in the expansion of each matrix element of $(\rho_A-\rho_A{(\infty}))$. As a result, the temporal fluctuation in the trace distance as the system relaxes is weaker when IPR is smaller. For states with high IPR, many $c_n$s will be close to $0$, which results in large temporal oscillations in trace distance from equilibrium. This causes a slower equilibration.

In Fig.~\ref{mpemba_timeevo_0p4}, for the \quotes{cooler} initial states, the initial distance from the diagonal ensemble is lower as compared to \quotes{hotter} states.  While we do not provide a rigorous proof, we offer a plausibility argument below. First, we consider the low IPR states. These states have a finite expectation value for Pauli observables. The diagonal ensemble constructed from this initial state will be a mixture of highly entangled mid-spectrum eigenstates.  However, the diagonal ensemble will have zero expectation value for such observables (as it is an almost uniform mixture of mid-spectrum eigenstates, which resemble infinite temperature states according to ETH). This suggests that the diagonal ensemble is not a good approximation at the initial time for the direct product state, resulting in a large initial trace distance. Note that states close to the ground state areapproximated by a diagonal ensemble of low-lying eigenstates that can have finite local expectation values.

 We present the IPR as a function of the initial-state parameters in Fig.~\ref{mpemba_timeevo_0p4}(a). Interestingly, the IPR shows a clear correlation with energy (Fig.~\ref{init_mpemba_energy}). The states with energy close to 0 (i.e, they are similar to mid-spectrum states) have a lower IPR (i.e., are more delocalized in the eigenspectrum). The states away from the middle of the spectrum have higher IPR (i.e., are more localized in the eigenspectrum). The states considered for time evolution are marked in the parameter space with dots. The states with  $\theta=0.5\pi$ show a spread in their IPR values.  The IPR distribution is consistent with the observation in Fig.~\ref{mpemba_timeevo_0p4}(d). The states with smaller IPR  decay faster despite having a larger initial trace distance from their equilibrium density matrix compared to states that have a larger IPR. A smaller IPR signifies a larger number of fluctuating phases in Eq.~\ref{tr_dist_mpemba}, which explains the faster relaxation. Further, we examine the overlap of the initial states used in Fig.~\ref{mpemba_timeevo_0p4}(b) with individual energy eigenstates and present the results in Fig.~\ref{mpemba_timeevo_0p4}(c). The state with energy close to $0$ has an almost uniform distribution of overlaps (with the eigenstates). The state away from zero energy show a larger concentration of overlap at the edge of the spectrum. The inset in Fig.~\ref{mpemba_timeevo_0p4}(c) shows the IPR for these states. In the states with $\theta=0.95\pi$, the IPR shows only a weak dependence on $\phi$ (similar to the behavior of the energy). This explains the absence of a clear QME for this set of states.

The clear observation of the QME in the chaotic regime is driven by the broad delocalization of initial states across the energy spectrum. A natural question is how this phenomenon behaves as the system approaches the integrable limit at high connectance ($\tilde{M} \to 1$), where the eigenspectrum deviates from chaotic WD statistics. We investigate this interplay at $\tilde{M}=0.8$ (Fig.~\ref{mpemba_timeevo_0p8}). While the correlation between IPR and energy persists (Fig.~\ref{init_mpemba_energy} and~\ref{mpemba_timeevo_0p8}(a) ), the overall range of IPR values is smaller, indicating that the initial states are more localized in the eigenbasis as $\tilde{M} \to 1$. Consequently, the nearly-polarized $\theta=0.95 \pi$ states still show no strong signature of the QME (Fig.~\ref{mpemba_timeevo_0p8}(b)). For the $\theta=0.5 \pi$ states, however, the dynamics are markedly different from the chaotic case. Instead of a smooth decay, the trace distance exhibits persistent late-time oscillations, as seen in Fig.~\ref{mpemba_timeevo_0p8}(d). We expect this to be caused by the emergence of symmetry constraints and (approximate) Hilbert space fragmentation close to an integrable point. In integrable models, quenching from the ground state results in coherent propagation of stable quasiparticles, which manifests as persistent oscillations and multiple crossings in the relaxation curves~\cite{Rylands2024MicroscopicOrigin, chalas2024multiple}. For a near-integrable system, our results show multiple crossings and oscillatory behavior in trace distance for a finite system within the timescale studied (Fig.~\ref{mpemba_timeevo_0p8}(d)).

\typeout{} 
\bibliography{refs.bib}	

\end{document}